\documentclass[12pt]{article}
\usepackage{amssymb}
\usepackage{epsf}
\usepackage{epsfig}
\newcommand{\lsim}{
\mathrel{\hbox{\rlap{\hbox{\lower4pt\hbox{$\sim$}}}\hbox{$<$}}}}

\newcommand{\gsim}{
\mathrel{\hbox{\rlap{\hbox{\lower4pt\hbox{$\sim$}}}\hbox{$>$}}}}

\newcommand{\gev}{\, {\rm GeV}}

\newcommand{\be}{\begin{equation}}
\newcommand{\ee}{\end{equation}}
\newcommand{\bi}{\begin{itemize}}
\newcommand{\ei}{\end{itemize}}
\newcommand{\ord}{{\cal O}}
\newcommand{\It}{\item}

\newcommand{\bdm}{\begin{displaymath}}
\newcommand{\edm}{\end{displaymath}}
\newcommand{\beqa}{\begin{eqnarray}}
\newcommand{\eeqa}{\end{eqnarray}}
\newcommand{\nonu}{\nonumber}

\begin{document}
\begin{titlepage}
\vspace*{-0.5truecm}

\begin{flushright}

TUM-HEP-549/04\\
hep-ph/0406048
\end{flushright}

\vspace*{0.3truecm}

\begin{center}
\boldmath
{\Large{\bf Early $SU(4)_{\rm PS} \otimes SU(2)_{L} \otimes
SU(2)_{R} \otimes SU(2)_{H}$   

\vspace{0.3truecm}

Unification of Quarks and Leptons}} 

\unboldmath
\end{center}

\vspace{0.5truecm}

\begin{center}
{\bf Andrzej J. Buras,${}^a$ P.Q. Hung,${}^b$ Ngoc-Khanh Tran,${}^b$ 
Anton Poschenrieder${}^a$ and Elmar Wyszomirski${}^{a}$}
 
\vspace{0.4truecm}

${}^a$ {\sl Physik Department, Technische Universit\"at M\"unchen,
D-85748 Garching, Germany}

\vspace{0.2truecm}

${}^b$ {\sl Dept. of Physics, University of Virginia, 382 McCormick Road\\
P. O. Box 400714, Charlottesville, Virginia 22904-4714, USA} 

\end{center}

\vspace{0.7cm}
\begin{abstract}
\vspace{0.2cm}\noindent
We discuss various aspects of the early petite unification (PUT) of 
quarks and leptons 
based on the gauge group
$G_{\rm PUT}=SU(4)_{\rm PS} \otimes SU(2)_{L} \otimes
SU(2)_{R} \otimes SU(2)_{H}$. This unification takes place at the 
scale $M=\ord (1-2~{\rm TeV})$ and gives the correct value of
$\sin^{2}\theta_{W}(M_{Z}^{2})$
without the violation of the upper bound on 
the $K_L\to\mu e$ rate and the limits on FCNC processes.
These properties require the existence of
three new generations of unconventional quarks and leptons
with charges up to $4/3$ (for quarks) and 2 (for leptons) 
and masses $\ord(250\gev)$
in addition to the standard three generations of quarks and leptons.
The horizontal group $SU(2)_H$  connects the standard
fermions with the unconventional ones.
We work out the spontaneous symmetry breaking (SSB) of the gauge group 
$G_{\rm PUT}$ down to the SM gauge group,
generalize the existing one-loop renormalization group (RG) analysis to the 
two-loop level including the contributions of Higgs scalars and Yukawa 
couplings, and
demonstrate that the presence of three new generations of heavy 
unconventional quarks and leptons with masses $\ord(250\gev)$ is consistent 
with astrophysical constraints. The NLO and Higgs contributions to the RG 
analysis are significant while the Yukawa contributions can be neglected.

\end{abstract}

\vspace*{0.5truecm}
\vfill
\noindent

\end{titlepage}

\thispagestyle{empty}
\vbox{}

\setcounter{page}{1}
\pagenumbering{roman}



\setcounter{page}{1}
\pagenumbering{arabic}

\section{Introduction}\label{sec:intro}
\setcounter{equation}{0}

The idea of Grand Unification (GUT) based on simple groups like SU(5) 
\cite{GG,BEGN} 
or SO(10) \cite{GG1,FM}, characterized by a single gauge coupling 
$g_{\rm GUT}$, is a 
very attractive scenario of 
the physics beyond the Standard Model (SM). In GUT models quarks 
and leptons are generally members of the same representation under the given 
gauge group, and this results in transitions that violate quark and lepton 
quantum numbers. As such transitions are very suppressed in nature, the GUT
scale must be very large, typically $\ord(10^{16} \gev)$, in 
order to be consistent with the experimental data, in 
particular with the lower bound on the proton life-time.

A less ambitious program is the Petite Unification \cite{HBB,BH03} that aims 
at 
unifying quarks 
and leptons at some energy scale $M$, not much greater than the electroweak 
scale, with 
the gauge group $G_S\otimes G_{W}$ that is characterized by two independent 
couplings $g_S$ and $g_W$. It is 
further assumed that $G_S$ and $G_W$ are either simple or pseudosimple 
(a direct product of simple groups with identical couplings). An attractive 
choice for the strong group $G_S$ 
is $SU(4)$ a la Pati-Salam \cite{PASA} with the lepton number playing the 
role of the
fourth colour. It turns out \cite{BH03} that with this choice of $G_S$ only very few 
weak groups $G_W$  can have low unification scale being 
simultaneously consistent with the measured value of 
$\sin^{2}\theta_{W}(M_{Z}^{2})$, the upper bound on the rare 
decay $K_L\to \mu e$ and the data on flavour changing neutral current (FCNC) 
processes. 
Basically only two weak gauge groups $SU(2)^3$ and $SU(3)^2$ can be made 
consistent with the experimental data. 

The general properties of the $SU(4)_{\rm PS}\otimes SU(2)^3$  and 
$SU(4)_{\rm PS}\otimes SU(3)^2$  unifications have been discussed in 
\cite{BH03}. In 
these models the values of $\sin^{2}\theta_{W}$
at the unification scale $M$ turn out to be $1/3$ and 
$3/8$, but a very fast renormalization group evolution allows 
to obtain 
correct $\sin^{2}\theta_{W}(M_{Z}^{2})$ with $M\approx 1~{\rm TeV}$ and 
$M\approx 3~{\rm TeV}$, respectively.

Concentrating on the unification based on the gauge group 
\be\label{PUT1}
G_{\rm PUT}=SU(4)_{\rm PS} \otimes SU(2)_{L} \otimes
SU(2)_{R} \otimes SU(2)_{H},
\ee
let us recall three most interesting properties of this model:
\begin{itemize}
\item
In addition to the standard three generations of quarks and leptons, new 
three generations of unconventional quarks and leptons
with charges up to $4/3$ (for quarks) and 2 (for leptons) 
and masses $\ord(250\gev)$ are automatically present.  
The horizontal group $SU(2)_H$  connects the standard
fermions with the unconventional ones.
\item
The placement of the ordinary quarks and leptons in the fundamental 
representation 
of $SU(4)_{\rm PS}$ is such that there
are {\em no tree-level} transitions between ordinary quarks and leptons
mediated by the $SU(4)_{\rm PS}$ gauge bosons. This prevents rare decays 
such as $K_L \rightarrow \mu e$ from acquiring
large rates, even when the masses of these gauge bosons are 
in the few TeV's range.
\item
There are new contributions to flavour changing neutral current
processes (FCNC) involving standard quarks and leptons that are mediated 
by the horizontal $SU(2)_H$ weak gauge bosons and the new
unconventional quarks and leptons. However, they appear first at the 
one--loop level and can be made consistent with the existing experimental 
bounds. 
\end{itemize}

It should be emphasized that the existence of new heavy fermions that 
are placed in the fundamental representation 
of $SU(4)_{\rm PS}$ together with ordinary fermions is essential for
having early unification of quarks and leptons without the problems 
with the rare decays like $K_L \rightarrow \mu e$. The original 
petite unification group $SU(4)_{\rm PS}\otimes SU(2)^4$ \cite{HBB}
having only ordinary fermions in the fundamental representation of 
$SU(4)_{\rm PS}$ and consequently $K_L \rightarrow \mu e$ proceeding 
at the tree level is ruled out as the early unification model unless further
new physics, such as large extra dimensions, is invoked \cite{CHP}.
In comparison with the usual
left--right symmetric models based on the
$SU(4)_{\rm PS}\otimes SU(2)_L\otimes SU(2)_R$ 
group \cite{LR1,LR2}, that is relevant for the grand $SO(10)$ unification, 
the present model 
has an additional $SU(2)_H$ factor and the fermion representations that 
differ from the latter case.
These new ingredients allow a low unification scale that is not possible in
the $SU(4)_{\rm PS}\otimes SU(2)_L\otimes SU(2)_R$ model.

In the present paper we would like to extend the analysis of the unification 
in (\ref{PUT1}) presented in \cite{BH03} by
\begin{itemize}
\item
working out the spontaneous symmetry breaking (SSB) of the gauge group 
$G_{\rm PUT}$ down to the SM gauge group,
\item
generalizing the one-loop renormalization group analysis of \cite{BH03} to the 
two-loop level and including the contributions of Higgs scalars and Yukawa 
couplings,
\item
demonstrating that the presence of three new generations of heavy 
unconventional quarks and leptons with masses $\ord(250\gev)$ is consistent 
with astrophysical constraints.
\end{itemize}
A short discussion of the rare decay $K_L\to\mu e$ and of FCNC processes
was already presented in \cite{BH03} and will be elaborated on elsewhere.

Our paper is organized as follows. In Section 2 we recall the main 
ingredients of the 
model in question, presenting in particular the fermion representations. 
In Section 3 
we present the Higgs system that accomplishes the desired SSB of 
$G_{\rm PUT}$  down to the SM 
group and work out the formulae for gauge boson and fermion masses. 
In Section 4 
we set up two-loop renormalization group equations for the evolution of the 
gauge couplings and determine the petite unification scale $M$ using as the 
inputs $\sin^{2}\theta_{W}(M_{Z}^{2})$, $\alpha_s(M_Z^2)$ and 
$\alpha(M_Z^2)$. 
 We present a very simple formula for $M$ as a function of these three inputs 
that to a very high accuracy reproduces our numerical analysis. 
In Section 5 we address the fate of the new heavy fermions in the context of
astrophysical constraints. Our conclusions and a brief outlook are given
in Section 6.

\section{The Model}
\setcounter{equation}{0}
\subsection{Preliminaries}
In this section we will describe the main ingredients of the model based on
the group $G_{\rm PUT}$ in (\ref{PUT1}). After presenting the pattern of the 
SSB and of the conditions for the coupling constants, we will describe in 
turn the gauge boson sector and the fermion sector.
The Higgs system responsible for the SSB will be presented in Section 3.
\subsection{General Structure}
The pattern of the SSB breaking is assumed to be
\begin{equation}
\label{pattern}
G_{\rm PUT} \stackrel{\textstyle M}{\longrightarrow} G_1 
\stackrel{\textstyle \tilde{M}}{\longrightarrow} G_2
\stackrel{\textstyle M_Z}{\longrightarrow} SU(3)_c \otimes U(1)_{EM} ,
\end{equation}
where
\begin{equation}
\label{G1}
G_1 = SU(3)_{c}(g_3) \otimes U(1)_S(\tilde{g}_S) \otimes
SU(2)_{L}(g_{2L})\otimes SU(2)_{R}(g_{2R})\otimes SU(2)_{H}(g_{2H})                        \, ,
\end {equation}
and
\begin{equation}
\label{G2}
G_2 = SU(3)_{c}(g_3) \otimes SU(2)_{L}(g_2) \otimes U(1)_{Y}(g_1)\,
\end {equation}
is the SM group.
In order to streamline the notation we will denote the usual hypercharge 
coupling $g^\prime$ 
by $g_1$ and $g_{\rm QCD}$ by $g_3$, reserving 
the index ``$S$" for the $U(1)_S$ group.
Moreover, if not specified, the coupling $g_2$ will always stand for $g_{2L}$.

Thus at some scale $M$, the strong $SU(4)_{\rm PS}$ group is broken down to
the product of the gauge symmetry group of QCD and the strong $U(1)_S$ group, 
corresponding to the unbroken diagonal generator $T_{15}$ of the 
$SU(4)_{\rm PS}$ group that does not belong to $SU(3)_c$. The explicit 
expression for $T_{15}$ is given in (\ref{T15}).

In the next step, at scale $\tilde M\le M$, the subgroup 
$U(1)_S\otimes SU(2)_{R}\otimes SU(2)_{H}$ of the $G_1$ group is broken down
to the weak hypercharge $U(1)_Y$ group. The generator $T_Y$ of $U(1)_Y$ 
is given by
\be\label{TY}
T_Y=C_R T_{3R}+C_H T_{3H}+C_S T_{15}
\ee
with $T_{3R}$ and $T_{3H}$ being the diagonal generators of 
$SU(2)_{R}$ and $SU(2)_{H}$, respectively.
Consequently, the electric charge generator is given by
\be
Q=T_{3L}+T_Y=Q_W+C_S T_{15}
\ee
where $Q_W$ is the "weak" charge corresponding to the group $G_W$.

The coefficients $C_i$ in (\ref{TY}) describe the embedding of
the weak hypercharge $U(1)_{Y}$
group into $G_1$. 
The fact that 
the weak hypercharge $U(1)_{Y}$
group merges into both $\tilde{G}_S$ and $G_W$ at $\tilde{M}$,
allows us to put quarks and leptons into identical
representations of the weak group $G_W$ and consequently make the quarks
and leptons to be indistinguishable when the strong interactions are turned 
off.
In the model in question, the coefficients $C_i$ are given by 
\be\label{CC}
C_W^2=C_R^2+C^2_H=2, \qquad  C^2_S=\frac{8}{3}.
\ee
These two values play an important role in the RG analysis presented in 
Section 4.
 
In particular, $C_W^2$ appears in the crucial group-theoretical
prefactor
\be
\sin^{2}\theta_{W}^0=
\frac{1}{1+C^2_W}=\frac{1}{3}~,
\ee
which, for the present discussion, is simply $1/3$. The reader is
urged to consult \cite{HBB} and \cite{BH03} for prefactors
corresponding to other choices of $G_W$.

Renormalization group effects in the range $M_Z\le \mu \le M$ can decrease 
 the mixing angle down to the experimental value
$\sin^{2}\theta_{W}(M_{Z}^{2})\approx 0.23$, provided the unification scale
and the representations of the matter fields (fermions and scalars) are
properly chosen. 
An explicit one--loop relation between $\sin^{2}\theta_{W}(M_{Z}^{2})$ 
and $\sin^{2}\theta_{W}^0$ is given by \cite{HBB,BH03}
\be
\label{sinsq2}
\sin^{2}\theta_{W}(M_{Z}^{2}) = \sin^{2}\theta_{W}^{0}[1-
C_{S}^{2}\frac{\alpha(M_{Z}^{2})}{\alpha_{3}(M_{Z}^{2})}-8\pi\cdot 
 \alpha(M_{Z}^{2})(K\ln\frac{\tilde{M}}{M_Z}+
K^{'}\ln\frac{{M}}{\tilde M})]\, ,
\ee
where
\be
\alpha(M_{Z}^{2}) \equiv \frac{e^{2}(M_{Z}^{2})}{4\pi}, \qquad
\alpha_{3}(M_{Z}^{2}) \equiv \frac{g_{3}^{2}(M_{Z}^{2})}{4\pi},
\ee
and the coefficients $K$ and $K^\prime$ are given in Section 4.

It has been demonstrated in \cite{BH03} that with the values of $C_W^2$ and
$C^2_S$ in (\ref{CC}) and the fermion representations specified below, the 
resulting values of $K$ and $K^\prime$, allow to obtain the correct 
value of $\sin^{2}\theta_{W}(M_{Z}^{2})$ provided 
$M\approx \tilde M \approx \ord(1~{\rm TeV})$. At the two--loop level that we
discuss in Section 4, the RG analysis is much more involved and we will 
proceed differently. We will use the experimental value of 
$\sin^{2}\theta_{W}(M_{Z}^{2})$ as an input to the evolution of the gauge 
couplings and will determine the unification scale $M$, in analogy to GUT 
analyses, by studying the ``matching" conditions for the relevant couplings. 
While these conditions will be spelled out systematically in Section 4, the 
basic relations behind them are \cite{BH03}
 
\begin{equation}
\label{e2}
\frac{1}{e^{2}(M_{Z}^{2})} =\frac{1}{[g_{2}(M_{Z}^{2})]^2} +
\frac{1}{[g_1(M_{Z}^{2})]^2}\, ,
\end{equation}

\begin{equation}
\label{g2}
g_{2L}(\tilde{M}^2)=g_{2R}(\tilde{M}^2)=g_{2H}(\tilde{M}^2)
=g_{W}(\tilde{M}^2)\, ,
\end{equation}

\begin{equation}
\label{e2p}
\frac{1}{[g_1(\tilde{M}^{2})]^2} =\frac{C^2_W}
{[g_{W}(\tilde{M}^{2})]^2} +\frac{C_{S}^2}
{[\tilde{g}_{S}(\tilde{M}^{2})]^2}\, ,
\end{equation}
with $g_W=g_2$ and

\be\label{conM}
g_{3}(M^2)= \tilde{g}_{S}(M^2) = g_{S}(M^2) \, .
\ee
Finally, we will use the standard $\overline{MS} $ definition for 
$\sin^{2}\theta_{W}$, namely
\begin{equation}
\label{sinsq}
\sin^{2}\theta_{W}(M_{Z}^{2}) = \frac{e^{2}(M_{Z}^{2})}
{g_{2L}^{2}(M_{Z}^{2})}\,.
\end{equation}

\subsection{Gauge Bosons}
\subsubsection{Gluons and Leptoquarks}
The $SU(3)_{c}$ content of the adjoint representation 
$\underline{15}$ of $SU(4)_{\rm PS}$ is as follows:
\begin{equation}
\underline{15} = \underline{8} + \underline{3} + 
\overline{\underline{3}} + \underline{1},
\end{equation}
and the corresponding gauge fields $A^{i}_{\mu S}$ are represented by

\begin{equation}
\left\{A^{i}_{\mu S}\left(i = 1, ..., 15\right)\right\} = 
\left(\begin{array}{cccc}   &   &   & G^{+, 1}_{\mu} \\   
& A^{i}_{\mu S}  &   &  \\   &   &   & G^{+, 2}_{\mu} \\   
& (i = 1, ..., 8) &   &  \\   &   &   & G^{+, 3}_{\mu}  \\   
&   &   &   \\ G^{-, 1}_{\mu} & G^{-, 2}_{\mu} 
& G^{-, 3}_{\mu} & \tilde{A}_{\mu S} \end{array}\right)
\end{equation}
Here the octet $A^{i}_{\mu S}$ stands for the gluons,

\begin{eqnarray}
G^{\pm, 1}_{\mu} &=& \frac{1}{\sqrt{2}}\left(A^{9}_{\mu S} 
\mp iA^{10}_{\mu S}\right),\\
G^{\pm, 2}_{\mu} &=& \frac{1}{\sqrt{2}}\left(A^{11}_{\mu S} 
\mp iA^{12}_{\mu S}\right),\\
G^{\pm, 3}_{\mu} &=& \frac{1}{\sqrt{2}}
\left(A^{13}_{\mu S}\mp iA^{14}_{\mu S}\right),
\end{eqnarray}
and $\tilde{A}_{\mu S} = \tilde{A}^{15}_{\mu S} $ is the neutral 
gauge boson which corresponds to the generator $T_{15}$ of $SU(4)_{S}$ 
and equivalently to the generator of $U(1)_{S}$. Explicitly
\be\label{T15}
T_{15}=\frac{1}{2\sqrt{6}}
\left(\begin{array}{cccc}
1   &   &   &  \\
    & 1 &   &   \\
    &   &  1 &   \\
    &   &    & -3 
\end{array}\right).
\end{equation}

The leptoquark gauge bosons $G^{\pm, i}_{\mu}$ carry electric 
charges $\pm \frac{4}{3}$ and connect quarks to leptons. 
They are responsible for the rare transitions, like $K_L\to \mu e$,
 the phenomenology of which 
 has been briefly presented in \cite{BH03} and will be presented in detail
elsewhere. Under the breaking of $SU(4)_{S}$ 
down to $SU(3)_{c} \times U(1)_{S} $ the leptoquarks $G^{\pm, i}_{\mu}$ 
gain masses of order $M$ whereas the gluons and the gauge boson 
$\tilde A_{\mu S}$ remain massless.

\subsubsection{Electroweak Gauge Bosons}

The model has nine massive electroweak gauge bosons in addition to 
the massless photon. These include (i) four charged gauge bosons 
$W^{\pm}_{R}, W^{\pm}_{H}$ and two neutral gauge bosons $Z_{1}, Z_{2}$ 
all with masses of order $\tilde{M}$, and (ii) the standard 
$W^{\pm}$ and $Z^{0}$ gauge bosons with the conventional masses that are 
very precisely measured. 
Below we summarize the structure of the neutral gauge boson sector. The 
derivation of these formulae and the discussion of the gauge boson masses is 
postponed to Section 3.

It should be remarked that the field $B_{\mu}$ of the SM is 
expressed in terms of the fields $\tilde{A}_{\mu S}$ and 
$\left(W^{3}_{\mu}\right)_{R, H}$, which couple to the diagonal 
generators of $U(1)_{S} \otimes SU(2)_{R} \otimes SU(2)_{H}$, 
respectively, as follows:

\begin{equation}\label{Bmu}
B_{\mu} = \cos\theta_{S} \tilde{A}_{\mu S} + 
\frac{\sin\theta_{S}}{\sqrt{2}}\left(W^{3}_{\mu R} + W^{3}_{\mu H}\right),
\end{equation}
where the mixing angle $\theta_{S}$ is defined by

\begin{equation}
\tan\theta_{S} = \frac{\tilde{g}_{S}}{g_{W}}\frac{\sqrt{3}}{2}.
\end{equation}
Furthermore, we have
\begin{eqnarray}\label{Z1Z2}
Z_{1\mu} &=& \sin\theta_{S} \tilde{A}_{\mu S}  - 
\frac{\cos\theta_{S}}{\sqrt{2}}\left(W^{3}_{\mu R} + W^{3}_{\mu H}\right),\\
Z_{2\mu} &=& \frac{1}{\sqrt{2}}\left(W^{3}_{\mu H} - W^{3}_{\mu R}\right). 
\end{eqnarray}

The masses of $Z_{1\mu}$ and $Z_{2\mu}$ are related through the relation 
\begin{equation}
M_{Z_{1}} = \frac{M_{Z_2}}{\cos\theta_{S}},\label{formula1}
\end{equation}
which will be derived in Section 3. 
Equation (\ref{formula1}) is the analog of the SM 
relation $M_{W} = M_{Z}\cos\theta_{W}$. $\tilde{M}$ 
must be larger than 800 GeV in order for the model to be consistent 
with the experimental data. Finally, it follows from (\ref{e2p}) 
that the hypercharge $U(1)_{Y}$ coupling constant $g_1$ is defined  
in terms of $\tilde{g}_{S}$ and $g_{W}$  by

\begin{equation}\label{egs}
g_1 = \frac{\sqrt{3}\tilde g_S g_{W}}
{\left(6\tilde g_S^2 + 
8 g^{2}_{W}\right)^{1/2}} = \frac{g_{W} \sin\theta_{S}}{\sqrt{2}}
\end{equation}
with all couplings evaluated at $\mu=\tilde M$.
This equation is analogous to the well-known relation $e=g\sin\theta_W$.

\subsection{Fermions}
The fermions in {\it each generation} can be divided into two groups 
(with the electric charges shown in parentheses):

\newpage

\noindent
{\bf a) Ordinary Fermions}

\be\label{SF}
\psi^{q}_{L,R} = \left(
\begin{array}{c}
u(2/3)\\ 
d(-1/3)
\end{array}
\right)_{L,R},
\qquad
\psi^{l}_{L,R} = \left(
\begin{array}{c}
\nu(0) \\ 
l(-1)
\end{array}
\right)_{L,R},
\ee
with $\psi^{q,l}_{L}$ and $\psi^{q,l}_{R}$ transforming as $(2,1)$ and
$(1,2)$  under 
$SU(2)_L\otimes SU(2)_R$, respectively.

\noindent
{\bf b) New Heavy Fermions}

\be\label{NHF}
\tilde{Q}_{L,R} = \left(
\begin{array}{c}
\tilde{U}(4/3)\\ 
\tilde{D}(1/3)
\end{array}
\right)_{L,R}, 
\qquad
\tilde{L}_{L,R} = \left(
\begin{array}{c}
\tilde{l}_{u}(-1)\\ 
\tilde{l}_{d}(-2)
\end{array}
\right)_{L,R},
\ee
with unconventional charges and masses of $\ord(M_F)$.
The left-handed and right-handed fields transform as
$(2,1)$ and $(1,2)$  under 
$SU(2)_L\otimes SU(2)_R$, respectively.
Thus their transformation properties under
$SU(2)_L\otimes SU(2)_R$ are the same as of ordinary fermions but due to 
different electric charges, they cannot be considered as new generations of 
ordinary quarks and leptons. The three generations of these new heavy fermions 
constitute for themselves a new set of sequential generations, that are 
connected with the ordinary generations through the interactions mediated by 
the $SU(2)_H$ gauge bosons and the leptoquarks $G^{\pm, i}_{\mu}$.

In \cite{BH03} also the third group of 
vector-like fermions
\be\label{VF}
\tilde{Q}^{\prime,\prime\prime}_{L,R} = \left(
\begin{array}{c}
\tilde{U}^{\prime,\prime\prime}(5/6)\\ 
\tilde{D}^{\prime,\prime\prime}(-1/6)
\end{array}
\right)_{L,R}
\qquad
\tilde{L}^{\prime,\prime\prime}_{L,R} = \left(
\begin{array}{c}
\tilde{l}^{\prime,\prime\prime}_{u}(-1/2)\\ 
\tilde{l}^{\prime,\prime\prime}_{d}(-3/2)
\end{array}
\right)_{L,R}
\ee
with unconventional charges and masses of $\ord(\tilde M)$ has been considered.
Here $\tilde{Q}^{\prime}_{L,R}$ and $\tilde{L}^{\prime}_{L,R}$ transform as 
doublets under $SU(2)_L$ and are singlets under $SU(2)_R\otimes SU(2)_H$, 
whereas $\tilde{Q}^{\prime\prime}_{L,R}$ and 
$\tilde{L}^{\prime\prime}_{L,R}$ transform as 
doublets under $SU(2)_R$ and are singlets under $SU(2)_L\otimes SU(2)_H$.
These additional fermions relevant in principle for the renormalization group
analysis for scales $\mu>\tilde M$ were introduced in \cite{BH03} to keep 
the $SU(2)_L$, $SU(2)_R$ and $SU(2)_H$ couplings equal under renormalization 
group evolution at least at the one loop level. Meanwhile we have realized 
that this equality is broken already at the one loop level by Higgs 
contributions and consequently there is really no reason to introduce them at
all. Moreover, as discussed in Section 5, due to very strange charges the 
existence of these fermions is problematic for cosmology. Therefore we 
will exclude them from our analysis.

Until now we have given only the representations under $SU(2)_L$ and
$SU(2)_R$. In order to avoid dangerous tree level transitions involving 
ordinary fermions that are mediated 
by the $SU(4)_{\rm PS}$ gauge bosons and FCNC transitions mediated by the 
``horizontal" $SU(2)_H$ gauge bosons, we proceed as follows:
\begin{itemize}
\item
With respect to $SU(4)_{\rm PS}$, we put ordinary quarks $\psi^{q}_{L,R}$ 
together with the heavy leptons $\tilde{L}_{L,R}$ into the fundamental 
representations of this group and similarly for  
ordinary leptons $\psi^{l}_{L,R}$ and heavy quarks $\tilde{Q}_{L,R}$.
\item
With respect to $SU(2)_{H}$, we put ordinary quarks $\psi^{q}_{L,R}$ 
together with heavy quarks $\tilde{Q}_{L,R}$ in the doublets of this group
and similarly for ordinary leptons $\psi^{l}_{L,R}$ and heavy 
leptons $\tilde{L}_{L,R}$. 
\end{itemize}

Explicitly then, the fermions are placed in the 
representations under $G_{\rm PUT}$ as follows:

\begin{equation}\label{REP1}
\Psi_L=(4,2,1,2)_L=
\left(\begin{array}{cc}
(d^c(1/3),\tilde U(4/3)) & (\tilde l_u(-1),\nu(0)) \\
(u^c(-2/3),\tilde D(1/3)) & (\tilde l_d(-2),l(-1))
\end{array}\right)_L
\end{equation}

\begin{equation}\label{REP2}
\Psi_R=(4,1,2,2)_R=
\left(\begin{array}{cc}
(d^c(1/3),\tilde U(4/3)) & (\tilde l_u(-1),\nu(0)) \\
(u^c(-2/3),\tilde D(1/3)) & (\tilde l_d(-2),l(-1))
\end{array}\right)_R
\end{equation}
where in order to put the ordinary quarks into representations of 
$G_{\rm PUT}$ we used their charge conjugated fields, suppressing a minus sign
in front of the $u^c$ field.
Each entry in the $2\times2$ matrices in (\ref{REP1}) and (\ref{REP2}) 
represents an $SU(2)_H$ doublet, whereas the columns represent 
$SU(2)_{L,R}$ doublets. With respect to $SU(4)_{\rm PS}$ we do not show 
explicitly the colour indices but $d^c(1/3)$ is placed in the quartet 
together with $\tilde l_u(-1)$, $\tilde U(4/3)$ together with $\nu(0)$ 
and analogously for fermions in  second rows in (\ref{REP1}) and 
(\ref{REP2}).

\section{The Spontaneous Symmetry Breaking}
\setcounter{equation}{0}
\subsection{\label{choice}Choices of Higgs Scalars}

In this section, we will discuss the Higgs sector which is needed to
spontaneously break $SU(4)_{\rm PS} \otimes SU(2)_{L} \otimes
SU(2)_{R} \otimes SU(2)_{H}$ down to the SM group. In particular, we will focus on
those scalars that can make an important contribution to the RG evolution
of the gauge couplings of the SM. 

The Higgs scalars that are needed are the following:

\begin{itemize}

\item The Higgs field that breaks $SU(4)_{PS}$ can be written as
\begin{equation}
\label{phis}
\Phi_{S} = (15,1,1,1)   \,.
\end{equation}
The vacuum expectation value (VEV) of
this Higgs field breaks $SU(4)_{\rm PS}$ down to $SU(3)_c \otimes
U(1)_S$ at a scale $M$.

\item Below $M$, one has effectively $SU(3)_c \otimes SU(2)_L
\otimes SU(2)_R \otimes SU(2)_H \otimes U(1)_S$. One
would like to find a scalar field that can spontaneously break
$SU(2)_L \otimes SU(2)_R \otimes SU(2)_H \otimes U(1)_S$ down
to $SU(2)_L \otimes U(1)_Y$ at some scale $\tilde{M} > M_Z$. 
Such a scalar should transform non-trivially under 
$SU(2)_R \otimes SU(2)_H \otimes U(1)_S$, in particular it should
have a \em{non vanishing} $U(1)_S$ \em{quantum number}. 
Furthermore, we require the $U(1)_Y$ gauge boson, $B_{\mu}$, to be
a linear combination of $W^3_{\mu R}$, $W^3_{\mu H}$, and $\tilde{A}_{\mu S}$, which implies
that there should be mixing between these three gauge bosons.
On first look, it appears that there are several possibilities.

One might consider, for example, the following Higgs Fields:
\begin{equation}
\label{delr}
\Delta^{\alpha}_{R} = (4, 1, 3, 1) \qquad
\Delta^{\alpha}_{H} = (4, 1, 1, 3)  \,,
\end{equation}
where $\alpha =1,..4$ denotes the $SU(4)_{\rm{PS}}$ index. 




For symmetry reasons one might want to include the following Higgs field:
$\Delta^{\alpha}_{L} = (4, 3, 1, 1)$. However, its VEV would break
$SU(2)_L$ and there exist severe experimental bounds on the contribution
of any Higgs triplet to the $\rho$ parameter (already at tree level). 
These bounds imply that its VEV would have to be less than a few percent of the
SM VEV. For convenience, we will assume that this VEV is identically zero.

\item Finally there are Higgs fields which would break $SU(2)_L \otimes U(1)_Y$
down to $U(1)_{EM}$. Furthermore, these Higgs fields should also couple
to fermions as in the SM so as to give these fermions a mass. 
The fermions in our model transform as:
$(4,2,1,2)_L$ and $(4,1,2,2)_R$. A mass term would transform as the bilinear
\begin{equation}
\overline{(4,2,1,2)}_L \otimes (4,1,2,2)_R = (1+15,2,2,1+3).
\end{equation}
Therefore, 
in principle, we could have the following Higgs fields:
$(15,2,2,1)$, $(15,2,2,3)$, $(1,2,2,1)$, $(1,2,2,3)$. Any one of these Higgs
fields can be a suitable candidate for the symmetry breaking. The
question one might ask is whether or not they are all necessary. To study
this question, let us first concentrate on the $SU(4)_{\rm{PS}}$-singlet
Higgs fields. We consider first
\begin{equation}
\label{phi}
\Phi = (1,2,2,1) \, .
\end{equation}
As we shall see explicitely below, the VEV of $\Phi$ gives 
\em{equal masses} to \em{all fermions} (quarks and leptons, including
both conventionally and unconventionally charged fermions). This
fact alone necessitates additional Higgs fields. Let us then look at
\begin{equation}
\label{phih}
\Phi_H = (1,2,2,3)   \,.
\end{equation}
The VEV of this $SU(4)_{\rm{PS}}$-singlet, $SU(2)_H$-triplet Higgs field
would split the mass scales of the conventional fermions (both
quarks and leptons) from the unconventional ones, as shown below.
This, however, does not split the masses of the quarks and the leptons.
For this to happen, we again need additional Higgs fields.

To split quark and lepton masses (for both conventional and
unconventional fermions), it appears sufficient to just use
a Higgs field:
\begin{equation}
\label{phi15}
\phi^{\beta} = (15, 2, 2, 1)  \,,
\end{equation}
where $\beta = 1,..,15$. As we shall see below, in the coupling of
$\phi^{\beta}$ to fermions, it is convenient to write
\begin{equation}
\label{phi15prime}
\phi_{s} = \phi^{\beta}\,\frac{\lambda_{\beta}}{2}   \,,
\end{equation}
where $\lambda_{\beta}/2$ are the generators of $SU(4)_{\rm{PS}}$. 
Notice that, under the QCD subgroup $SU(3)_c$, a $15$ splits into
$8+3+\bar{3}+1$. Therefore it should be the $SU(3)_c$ singlet
part which develops a VEV. A VEV of
the form $\langle\phi_{s}\rangle = \langle \phi^{15} \rangle
\lambda_{15}/2$ would split the masses
of the quarks from those of the leptons. The origin of further splitting in the quark sector, in particular between the top quark and the other standard quarks will be discussed in Section 3.3.  
\end{itemize}

Finally, for symmetry reasons, one might also have
$(1,2,1,2)$, $(1,1,2,2)$ and $(15,2,1,2)$ Higgs representations. 
However, these scalars {\em do not couple} to fermions. The
electric and hypercharge structures of these Higgs fields
are identical to those of the representations $(1,2,2,1)$,
the color-singlet part of $(15,1,2,2)$,
and $(15,2,2,1)$ respectively and we will concentrate on the latter.

\subsection{\label{charges}Electric Charges of Selected Higgs Fields and their
Couplings to SM Gauge Bosons}

Since we are primarily interested in this paper in the contributions
of the scalar sector to the RG evolution of the SM gauge couplings
up to $M$,
we shall not discuss the case of $\Phi_S = (15,1,1,1)$.
We shall instead concentrate on $\Delta_{R,H}, \Phi,\,\Phi_{H},
\phi_{s}$. All Higgs fields in this paper are
{\em complex}.

As we have seen in \cite{HBB,BH03} and above, the $U(1)_S$ charge matrix
for $SU(4)_{PS}$ fundamental representations  is
\be
\label{hyperS}
Y_S = C_S T_{15}  
=\left(
\begin{array}{cccc}
1/3&0&0&0 \\
0&1/3&0&0 \\
0&0&1/3&0 \\
0&0&0&-1 \\
\end{array}
\right) \,.
\ee
We now use (\ref{hyperS}) to find the electric charge assignments
and the $U(1)_Y$ quantum numbers for the Higgs fields listed
above.

\begin{itemize}

\item For both $\Delta_{R}$ and $\Delta_{H}$, one has 
\begin{equation}
\label{Q3}
Q_W = (1,0,-1)   \,,
\end{equation}
since they are both triplets under their respective $G_W$ group.

With $Q=Q_W + Y_S$, one obtains the following electric charge
assignments for both $\Delta_{R,H}$:
\begin{equation}
\label{chargedel}
Q_{\Delta} = \{ (4/3,1/3,-2/3),\, (0, -1, -2)\}  \,,
\end{equation}
where the first entries inside the parentheses are for the color triplets and the second entries are for the color singlets.

One can now explicitely write $\Delta_{R}$ and $\Delta_{H}$. One
has:
\begin{equation}
\label{DelRH}
\Delta_{R,H}=\{(\Delta_{R,H}^{(4/3),i}, \Delta_{R,H}^{(1/3),i},
\Delta_{R,H}^{(-2/3),i}),(\Delta_{R,H}^{0}, \Delta_{R,H}^{-},
\Delta_{R,H}^{--})\}   \,,
\end{equation}
where $i=1,2,3$ is the color index.

Since $\Delta_{R}$ and $\Delta_{H}$ are $SU(2)_L$ singlets, the $U(1)_Y$
quantum numbers $Y$ are identical to $Q_{\Delta}$, namely
\begin{equation}
\label{Yhyper}
Y \equiv Q_{\Delta}   \,.
\end{equation}

In the discussion of symmetry breaking given below, it is convenient
to express $\Delta_{R}$ and $\Delta_{H}$ as $2 \times 2$ matrices,
namely
\begin{equation}
\label{DelRH2}
\Delta_{R,H}^{S} \equiv \vec{\Delta}_{R,H}.\vec{T}_{R,H} =
\left(
\begin{array}{cc}
\Delta_{R,H}^{-}/2& \Delta_{R,H}^{0}/\sqrt{2}\\
\Delta_{R,H}^{--}/\sqrt{2}&-\Delta_{R,H}^{-}/2\\
\end{array}
\right) \,,
\end{equation}
for the {\em color singlet} part, and
\begin{equation}
\label{DelRH3}
\Delta_{R,H}^{i} \equiv \vec{\Delta}_{R,H}^{i}.\vec{T}_{R,H} =
\left(
\begin{array}{cc}
\Delta_{R,H}^{(1/3),i}/2& \Delta_{R,H}^{(4/3),i}/\sqrt{2}\\
\Delta_{R,H}^{(-2/3),i}/\sqrt{2}&-\Delta_{R,H}^{(1/3),i}/2\\
\end{array}
\right) \,,
\end{equation}
for the {\em color triplet} part. The superscript $S$ in 
 (\ref{DelRH2}) is a notation for ``color-singlet''.
For symmetry breaking, only the neutral scalar in (\ref{DelRH2})
can develop a VEV.

\item For $\Phi = (1,2,2,1)$, it is clear that one now has {\em two} $SU(2)_L$
doublets. Therefore each doublet interacts with the $SU(2)_L$ gauge
bosons in a standard way. As we have already shown in \cite{BH03},
the charge assignment for a representation $(1,2,2,1)$ is simply
\begin{equation}
\label{Phi}
Q \equiv Q_W = \left(
\begin{array}{cc}
0&1 \\
-1&0\\
\end{array}
\right) \,,
\end{equation}
where the first row (column) refers to $T_{3L}= 1/2$ ($T_{3R}= -1/2$)
and the second row (column) refers to $T_{3L}= -1/2$ ($T_{3R}= 1/2$).

One can explicitely write $\Phi$ in terms of a $2 \times 2$ matrix as
\begin{equation}
\label{Phi2}
\Phi = \left(
\begin{array}{cc}
\Phi_{1}^{0}&\Phi_{2}^{+} \\
\Phi_{1}^{-}&\Phi_{2}^{0}\\
\end{array}
\right) \,.
\end{equation}
The $U(1)_Y$ quantum number is in this case  simply 
\begin{equation}
\label{YPhi}
Y \equiv T_{3R} = \pm \frac{1}{2}  \,,
\end{equation}
where $Y=-1/2$ for the $SU(2)_L$ doublet with charge $(0,-1)$ and
$Y=+1/2$ for the $SU(2)_L$ doublet with charge $(1,0)$.

\item The charge assignment for $\Phi_H = (1,2,2,3)$ is just 
slightly more complicated. It is simply a direct
sum of $Q_W$ (\ref{Phi}) and $T_{3H} = (1,0,-1)$ (for a triplet),
namely
\begin{equation}
\label{PhiH}
Q = Q_W \oplus T_{3H}   \,.
\end{equation}
 From (\ref{PhiH}), one obtains the following $2\times 2$ charge matrices
referring to $(2,2)$ under $SU(2)_L \otimes SU(2)_R$ and there are {\em
three} of those (with the $U(1)_Y$ quantum numbers listed next to them):

\begin{equation}
\label{PhiH1}
\left(
\begin{array}{cc}
1&2 \\
0&1\\
\end{array}
\right) \,; Y=(\frac{1}{2}, +\frac{3}{2})  \,, 
\end{equation}

\begin{equation}
\label{PhiH2}
\left(
\begin{array}{cc}
0&1 \\
-1&0\\
\end{array}
\right) \,; Y=(-\frac{1}{2}, +\frac{1}{2})  \,,
\end{equation}

\begin{equation}
\label{PhiH3}
\left(
\begin{array}{cc}
-1&0 \\
-2&-1\\
\end{array}
\right) \,; Y=(-\frac{3}{2}, -\frac{1}{2})  \,.
\end{equation}

In these equations, the $U(1)_Y$
hypercharges $Y$ refer to the first and second column respectively.
Also equations (\ref{PhiH1}), (\ref{PhiH2}), (\ref{PhiH3}) refer to
{\em six} $SU(2)_L$ doublets which couple to the corresponding
gauge bosons in a standard way. The couplings to the $U(1)_Y$
gauge boson are given in terms of the hypercharges listed above.

Finally, one can write explicitely $\Phi_H$ as

\begin{equation}
\label{PhiH+}
\Phi_{H,+} = \left(
\begin{array}{cc}
\Phi_{H+,1}^{+}&\Phi_{H+,2}^{++} \\
\Phi_{H+,1}^{0}&\Phi_{H+,2}^{+}\\
\end{array}
\right) \,,
\end{equation}

\begin{equation}
\label{PhiH0}
\Phi_{H,0} = \left(
\begin{array}{cc}
\Phi_{H0,1}^{0}&\Phi_{H0,2}^{+} \\
\Phi_{H0,1}^{-}&\Phi_{H0,2}^{0}\\
\end{array}
\right) \,,
\end{equation}

\begin{equation}
\label{PhiH-}
\Phi_{H,-} = \left(
\begin{array}{cc}
\Phi_{H-,1}^{-}&\Phi_{H-,2}^{0} \\
\Phi_{H-,1}^{--}&\Phi_{H-,2}^{-}\\
\end{array}
\right) \,.
\end{equation}

The subscripts $+,0,-$ in the equations above refer to the three
scalars with $T_{3H} = (1,0,-1)$ respectively.

\item The Higgs field, whose VEV could split the
masses of the quarks from those of the leptons, could be
$\phi^{\beta}= (15,2,2,1)$ as mentioned above. To find
the charge structure of this Higgs field, it is useful to
first split it into $SU(3)_c$ representations:

1) {\bf{8}}: $\phi^{i}$ ($i=1,..,8$),

2) {\bf{3},${\bf{\bar{3}}}$}: $\phi^{\pm,1}=\frac{1}{\sqrt{2}}(\phi^{9} \mp i \phi^{10})$,
$\phi^{\pm,2}=\frac{1}{\sqrt{2}}(\phi^{11} \mp i \phi^{12})$,
$\phi^{\pm,3}=\frac{1}{\sqrt{2}}(\phi^{13}\mp i \phi^{14})$,

3) {\bf{1}}: $\phi^{15}$.

Notice that {\em each} one of the $\phi$'s written above is a $(2,2)$ under
$SU(2)_L \otimes SU(2)_R$.

The $U(1)_S$ quantum numbers, $Y_S$, of $\phi^{\beta}$ can easily be found.
They are exactly {\em the same} as the electric charges of the $SU(4)_{\rm{PS}}$ gauge bosons for the latter are singlets under $G_W$. Therefore:

1) $Y_S = 0$ for {\bf{8}}, and {\bf{1}},

2) $Y_S = \pm \frac{4}{3}$ for {\bf{3},${\bf{\bar{3}}}$}.

Since $Q= Q_W \oplus Y_S$, one obtains:

\begin{equation}
\label{phi80}
\left(
\begin{array}{cc}
0&1 \\
-1&0\\
\end{array}
\right) \,; Y=(-\frac{1}{2}, +\frac{1}{2}) \,: {\bf{8,1}}  \,,
\end{equation}

\begin{equation}
\label{phi3}
\left(
\begin{array}{cc}
4/3&7/3 \\
1/3&4/3\\
\end{array}
\right) \,; Y=(\frac{5}{6}, \frac{11}{6}) \,: {\bf{3}}  \,,
\end{equation}

\begin{equation}
\label{phi3bar}
\left(
\begin{array}{cc}
-4/3&-1/3 \\
-7/3&-4/3\\
\end{array}
\right) \,; Y=(-\frac{11}{6}, -\frac{5}{6}) \,: {\bf{\bar{3}}}  \,.
\end{equation}

Explicitely, we can write $\phi^{\beta}$ as

\begin{equation}
\label{phi8}
\phi^{(8)} = \left(
\begin{array}{cc}
\phi_{1}^{(8),0}&\phi_{2}^{(8),+} \\
\phi_{1}^{(8),-}&\phi_{2}^{(8),0}\\
\end{array}
\right) \,,
\end{equation}

\begin{equation}
\label{phi1}
\phi^{(1)} = \left(
\begin{array}{cc}
\phi_{1}^{(1),0}&\phi_{2}^{(1),+} \\
\phi_{1}^{(1),-}&\phi_{2}^{(1),0}\\
\end{array}
\right) \,,
\end{equation}

\begin{equation}
\label{phi3b}
\phi^{(3)} = \left(
\begin{array}{cc}
\phi_{1}^{(3),4/3}&\phi_{2}^{(3),7/3} \\
\phi_{1}^{(3),1/3}&\phi_{2}^{(3),4/3}\\
\end{array}
\right) \,,
\end{equation}

\begin{equation}
\label{phi3barb}
\phi^{(\bar{3})} = \left(
\begin{array}{cc}
\phi_{1}^{(\bar{3}),-4/3}&\phi_{2}^{(\bar{3}),-1/3} \\
\phi_{1}^{(\bar{3}),-7/3}&\phi_{2}^{(\bar{3}),-4/3}\\
\end{array}
\right) \,.
\end{equation}

\end{itemize}

 From the above listing of the $SU(2)_L$ doublets and their hypercharge
quantum numbers, one can proceed to include their contributions to
the evolution of the SM gauge couplings at one loop. At two loops,
in order to properly include the scalar contributions, one has to
work out the Yukawa couplings between some of the Higgs fields listed above
and the fermions.

Finally, as we mentioned above, in principle there could exist, for
symmetry reasons, the following Higgs fields: $(1,2,1,2)$,
$(1,1,2,2)$, and
$(15,2,1,2)$. The electric charge and hypercharge assignments for
these Higgs fields are identical to those for $\Phi$,
the {\bf 1} of $\tilde{\phi}^{\beta}$, and $\phi^{\beta}$
respectively.

\subsection{Yukawa Couplings}\label{yukawacouplings}

In this section, we will not make any serious attempt to construct
a model for fermion masses, but rather we are more interested in
a rough value for the Yukawa couplings as deduced from the overall
mass scales of the conventional and unconventional quarks and leptons.
This exercise serves as an estimate of the contributions of the 
Yukawa couplings to the two-loop beta functions.
For convenience, we shall make two assumptions concerning the
masses of the conventional and unconventional fermions. We believe
that our estimates of the unification mass scales will not be much
affected by details of fermion mass models. These assumptions
are as follows.

1) What we have in mind for the discussion that follows is some kind
of democratic-type mass matrices for the ordinary quarks
\cite{branco}. This implies that
the masses for the Up and Down sectors given below will be taken
to be universal mass scales which appear in front of $3 \times 3$
matrices whose elements have magnitudes of order unity. For this reason,
models of this type are usually referred to as Universal Strength for Yukawa
couplings (USY) because these couplings are common to all families
for each Up and Down sector. The hierarchy in masses comes from
the diagonalization of this type of matrices. The above ansatz
could be realized in a number of ways. One approach is
to go to more than three spatial dimensions as has been done
in \cite{hungmarcos} where it was found that the quark mass matrices
were of the almost-pure phase type (a generalized version of democratic
matrices). In \cite{hms}, this type of mass matrices was found to fit
the mass spectrum and the CKM matrix rather well. 

2) For the unconventional fermions, we have to keep in mind that
they are as yet unobserved. As a result, their masses are constrained
in various ways which, in general, depend on their lifetimes and
decay modes. If, for the sake of making some estimates as to their
contributions to the RG evolution, we assume that their masses are 
all equal to or larger than $250\gev$, then we are faced with
a very different mass pattern for these unconventional fermions.
The $3 \times 3$ matrices mentioned above for the ordinary quarks
would have to be very different for the unconventional ones. In fact,
most likely they would have to be of a form as to yield eigenvalues
of the same order so as to generate masses which would differ from each
other by at most a factor of two.

With the above two assumptions, the discussion which follows
deals uniquely with universal mass scales.

The fermions of our model are for each generation as follows:
\begin{equation}
\label{psil}
\Psi_L = (4,2,1,2)_L,\qquad
\Psi_R = (4,1,2,2)_R  \,.
\end{equation}

As we have discussed at length in \cite{BH03} and in Section 2, these
representations contain conventionally and unconventionally charged
quarks and leptons. As alluded to above, one needs to split the masses
of the conventional fermions from the unconventional ones since the
latter have not been observed experimentally. This can be achieved
by the use of both $\Phi$ and $\Phi_H$.

Then, the quark and leptonic $SU(2)_H$ doublets are

\begin{equation}
\label{quark}
(i\tau_{2} \psi^{q,*}_{L,R},\, \tilde{Q}_{L,R}),  \qquad
(\tilde{L}_{L,R},\, \psi^{l}_{L,R})  \,.
\end{equation}
We emphasize again that the way the fermions are ordered here
implies that there is {\em no} tree-level
transition between normal quarks and normal leptons due to the $SU(4)_{\rm{PS}}$ gauge bosons which link only conventional to unconventional fermions.

With (\ref{quark}) in mind one can now try to
write down the various Yukawa couplings. First, it is useful to
note the following identity:

\begin{equation}
\label{iden}
\overline{(i\tau_{2} \psi^{q,*}_{L})}\,(i\tau_{2} \psi^{q,*}_{R})=
(\bar{\psi^{q}}_L\, \psi^{q}_R)^{*}  \,.
\end{equation}
Since $\Phi_H$ is an $SU(2)_H$ triplet, it is convenient to write

\begin{equation}
\label{triplet}
\vec{\Phi}_{H} . \vec{T}_H =\left(
\begin{array}{cc}
\Phi_{H}^{3}/2&\Phi_{H}^{``+''}/\sqrt{2} \\
\Phi_{H}^{``-''}/\sqrt{2}&-\Phi_{H}^{3}/2\\
\end{array}
\right)  \,,
\end{equation}
where $\Phi_{H}^{``\pm''}= (\Phi_{H}^{1} \mp i \Phi_{H}^{2})/\sqrt{2}$.

Using  (\ref{iden}), one obtains for the Yukawa coupling to $\Phi$:

\begin{eqnarray}
\label{yukphi}
g_{\Phi} \bar{\Psi}_L \Phi \Psi_R + h.c.
&=& g_{\Phi} \{(\bar{\psi^{q}}_L\,\Phi^{*} \psi^{q}_R)^{*} +
\overline{\tilde{Q}_{L}}\Phi \tilde{Q}_{R}  \nonumber \\
& &+ (quarks \rightarrow leptons)\} + h.c. \,,
\end{eqnarray}
where $g_{\Phi}$ is taken to be positive.

Similarly, the Yukawa coupling to $\Phi_H$ can be written as
\begin{eqnarray}
\label{yukphih}
-g_{\Phi_H} \bar{\Psi}_L \vec{\Phi}_{H} . \vec{T}_H \Psi_R + h.c.
&=& g_{\Phi_H}\{\overline{\tilde{Q}_{L}}\frac{\Phi_{H}^{3}}{2} \tilde{Q}_{R}
-(\bar{\psi^{q}}_L\,\frac{\Phi_{H}^{3*}}{2} \psi^{q}_R)^{*} + (quarks 
\nonumber \\
& & \rightarrow leptons)+(charge-changing \nonumber \\ 
& &  interactions) + h.c.\} \,,
\end{eqnarray}
where $g_{\Phi_H}$ is taken to be positive.

When $\Phi$ and $\Phi_{H}^{3}$ develop a VEV, (\ref{yukphi}) and
\ref{yukphih}) give contributions to fermion
masses. First let us remind ourselves that both $\Phi$ and $\Phi_{H}^{3}$
are $2 \times 2$ matrices with respect to $SU(2)_L \otimes SU(2)_R$. Therefore
we have a different VEV for the Up and Down quark sector (as well as 
for its leptonic counterpart):

\begin{equation}
\label{VEVphi}
\left(
\begin{array}{cc}
\langle \Phi \rangle_{u}&0 \\
0& \langle \Phi \rangle_{d}\\
\end{array}
\right) \,
\end{equation}

\begin{equation}
\label{VEVphih}
\left(
\begin{array}{cc}
\langle \Phi_{H}^{3} \rangle_{u}&0 \\
0& \langle \Phi_{H}^{3} \rangle_{d}\\
\end{array}
\right) \,
\end{equation}

We list below various contributions to fermion masses.

I) Masses coming from $\Phi$ and $\Phi_H$.

\begin{itemize}

\item The Up quark sector:

\begin{equation}
\label{Upq}
m_{U}(\Phi, \Phi_{H}) = g_{\Phi}\langle \Phi \rangle_{u} -
g_{\Phi_H} \langle \Phi_{H}^{3} \rangle_{u}  \,,
\end{equation}

\begin{equation}
\label{Upqtil}
m_{\tilde{U}}(\Phi, \Phi_{H}) = g_{\Phi}\langle \Phi \rangle_{u} +
g_{\Phi_H} \langle \Phi_{H}^{3} \rangle_{u}  \,.
\end{equation}

\item The Down quark sector:

\begin{equation}
\label{Downq}
m_{D}(\Phi, \Phi_{H}) = g_{\Phi}\langle \Phi \rangle_{d} -
g_{\Phi_H} \langle \Phi_{H}^{3} \rangle_{d}  \,,
\end{equation}

\begin{equation}
\label{Downqtil}
m_{\tilde{D}}(\Phi, \Phi_{H}) = g_{\Phi}\langle \Phi \rangle_{d} +
g_{\Phi_H} \langle \Phi_{H}^{3} \rangle_{d}  \,.
\end{equation}

For reasons which are given below, we will (by a suitable definition
of the phase) define the lepton masses with a negative sign ( which
{\em cannot} be detected in any case). We have the following masses.

\item The Up lepton sector:

\begin{equation}
\label{Uplu}
-m_{\tilde{L}_u}(\Phi, \Phi_{H}) = g_{\Phi}\langle \Phi \rangle_{u} -
g_{\Phi_H} \langle \Phi_{H}^{3} \rangle_{u}  \,,
\end{equation}

\begin{equation}
\label{UpN}
-m_{N}(\Phi, \Phi_{H}) = g_{\Phi}\langle \Phi \rangle_{u} +
g_{\Phi_H} \langle \Phi_{H}^{3} \rangle_{u}  \,.
\end{equation}

\item The Down lepton sector:

\begin{equation}
\label{Downld}
-m_{\tilde{L}_d}(\Phi, \Phi_{H}) = g_{\Phi}\langle \Phi \rangle_{d} -
g_{\Phi_H} \langle \Phi_{H}^{3} \rangle_{d}  \,,
\end{equation}

\begin{equation}
\label{DownL}
-m_{L}(\Phi, \Phi_{H}) = g_{\Phi}\langle \Phi \rangle_{d} +
g_{\Phi_H} \langle \Phi_{H}^{3} \rangle_{d}  \,.
\end{equation}

\end{itemize}

II) Quark-lepton mass splitting from $\phi^{\beta}=(15,2,2,1)$.

As we have mentioned above, the component of $\phi^{\beta}=(15,2,2,1)$
which can develop a VEV is the color singlet $\phi^{(1)}$,
most specifically $\phi_{1}^{(1),0}$ and $\phi_{2}^{(1),0}$ in (\ref{phi1}).
Since $\phi^{(1)} \equiv \phi^{15} $, it follows that

\begin{equation}
\label{VEV15}
\langle \phi^{\beta}T_{\beta} \rangle =
\langle \phi^{15} \rangle T_{15}  \,,
\end{equation}
where diagonal matrix $T_{15}$ is given in (\ref{T15}).
Let us denote the Yukawa coupling of the fermions to $\phi^{\beta}$
by $g_{\phi}$. 

The total contribution of the fermion masses is now
given by the following expressions. Notice that we only give
the expression for the neutral lepton sector as Dirac masses. No 
attempt will be made in this paper concerning possible nature of the
neutrino mass. This will be dealt with in a subsequent paper.

\begin{itemize}

\item The Up quark and lepton sector:

\begin{equation}
\label{Upq2}
m_{U} = g_{\Phi}\langle \Phi \rangle_{u} -
g_{\Phi_H} \langle \Phi_{H}^{3} \rangle_{u} 
+ g_{\phi} \langle \phi_{1}^{(1),0} \rangle\,,
\end{equation}

\begin{equation}
\label{Upqtil2}
m_{\tilde{U}} = g_{\Phi}\langle \Phi \rangle_{u} +
g_{\Phi_H} \langle \Phi_{H}^{3} \rangle_{u} 
+ g_{\phi} \langle \phi_{1}^{(1),0} \rangle \,,
\end{equation}

\begin{equation}
\label{Uplu2}
-m_{\tilde{L}_u} = g_{\Phi}\langle \Phi \rangle_{u} -
g_{\Phi_H} \langle \Phi_{H}^{3} \rangle_{u}  
- 3g_{\phi} \langle \phi_{1}^{(1),0} \rangle\,,
\end{equation}

\begin{equation}
\label{UpN2}
-m_{N} = g_{\Phi}\langle \Phi \rangle_{u} +
g_{\Phi_H} \langle \Phi_{H}^{3} \rangle_{u}  
- 3g_{\phi} \langle \phi_{1}^{(1),0} \rangle \,.
\end{equation}

\item The Down quark and lepton sector:

\begin{equation}
\label{Downq2}
m_{D} = g_{\Phi}\langle \Phi \rangle_{d} -
g_{\Phi_H} \langle \Phi_{H}^{3} \rangle_{d} 
+ g_{\phi} \langle \phi_{2}^{(1),0} \rangle \,,
\end{equation}

\begin{equation}
\label{Downqtil2}
m_{\tilde{D}} = g_{\Phi}\langle \Phi \rangle_{d} +
g_{\Phi_H} \langle \Phi_{H}^{3} \rangle_{d}  
+ g_{\phi} \langle \phi_{2}^{(1),0} \rangle \,,
\end{equation}

\begin{equation}
\label{Downld2}
-m_{\tilde{L}_d} = g_{\Phi}\langle \Phi \rangle_{d} -
g_{\Phi_H} \langle \Phi_{H}^{3} \rangle_{d} 
- 3g_{\phi} \langle \phi_{2}^{(1),0} \rangle  \,,
\end{equation}

\begin{equation}
\label{DownL2}
-m_{L} = g_{\Phi}\langle \Phi \rangle_{d} +
g_{\Phi_H} \langle \Phi_{H}^{3} \rangle_{d} 
- 3g_{\phi} \langle \phi_{2}^{(1),0} \rangle  \,.
\end{equation}

\end{itemize}
 From (\ref{Upq2}), (\ref{Upqtil2}), (\ref{Uplu2}), one obtains the following products of the Yukawa couplings with the VEVs:

\begin{equation}
\label{gPhiu}
g_{\Phi}\langle \Phi \rangle_{u} = \frac{1}{4}(m_{U} + 2m_{\tilde{U}}-
m_{\tilde{L}_u}) \,,
\end{equation}

\begin{equation}
\label{gPhiHu}
g_{\Phi_H} \langle \Phi_{H}^{3} \rangle_{u} =
\frac{1}{4}(2m_{\tilde{U}} - 2m_{U})  \,,
\end{equation}

\begin{equation}
\label{gphi1}
g_{\phi} \langle \phi_{1}^{(1),0} \rangle =
\frac{1}{4}(m_{\tilde{L}_u} + m_{U})  \,,
\end{equation}

\begin{equation}
\label{gPhid}
g_{\Phi}\langle \Phi \rangle_{d} = \frac{1}{4}(m_{D} + 2m_{\tilde{D}}-
m_{\tilde{L}_d}) \,,
\end{equation}

\begin{equation}
\label{gPhiHd}
g_{\Phi_H} \langle \Phi_{H}^{3} \rangle_{d} =
\frac{1}{4}(2m_{\tilde{D}} - 2m_{D})  \,,
\end{equation}

\begin{equation}
\label{gphi2}
g_{\phi} \langle \phi_{2}^{(1),0} \rangle =
\frac{1}{4}(m_{\tilde{L}_d} + m_{D})  \,.
\end{equation}
From the above equations, one can estimate the size of various Yukawa couplings
by making educated guesses on the masses of various unconventional fermions
as well as the values of various VEVs.


%
%
%
%

\boldmath
\subsection{The Neutral Gauge Bosons from the Breaking
$SU(2)_R \otimes SU(2)_H \otimes U(1)_S \rightarrow U(1)_Y$}
\unboldmath
The main goal of this subsection is to find the expression
for the $U(1)_Y$ gauge boson $B_{\mu}$ in terms of the neutral gauge bosons
of $SU(2)_R \otimes SU(2)_H \otimes U(1)_S$. However, we will
also present similar expressions for the other two massive neutral gauge
bosons which accompany the massless $B_{\mu}$ at this stage of 
symmetry breaking.
We will postpone the discussion of mixings between charged gauge
bosons to Section 5 where we discuss the fate of the
unconventional fermions.

Before discussing how the neutral gauge bosons obtain their masses
through spontaneous symmetry breaking, let us explain how (\ref{Bmu})
was derived. From (\ref{TY}) for $T_Y$, one 
can readily write down an expression for the $U(1)_Y$ gauge field as

\begin{equation}
\label{Bmu1}
B_{\mu} = \frac{g_W}{\sqrt{g_W^2 C_S^2 + \tilde{g}_{S}^2 C_{W}^2}}\,C_{S}
\tilde{A}_{\mu S} + \frac{\tilde{g}_S}{\sqrt{g_W^2 C_S^2 + \tilde{g}_{S}^2 C_{W}^2}}
\sum_{i \neq SU(2)_L} C_{i\,W} W_{\mu i} \,.
\end{equation}
Putting into (\ref{Bmu1}) the explicit values $C_S^2 = 8/3$, $C_{W}^2=2$, one obtains for the coefficients in the notation of (\ref{Bmu}) explicitely

\begin{equation}
\label{thetas}
\cos \theta_S = \frac{2\,g_W}{\sqrt{3\,\tilde{g}_S^2 + 4\,g_W^2}},\qquad
\sin \theta_S = \frac{\sqrt{3}\,\tilde{g}_S}{\sqrt{3\,\tilde{g}_S^2 +4\, g_W^2}} \,.
\end{equation}
Furthermore, from (\ref{e2p}), it is straightforward to derive (\ref{egs}).

Let us now see how one obtains (\ref{Bmu}) and the two massive neutral gauge bosons by explicit coupling to the Higgs fields.

In this discussion, it is convenient to write the $SU(2)_{R,H}$ gauge bosons in terms of a $2 \times 2$ matrix as follows:

\begin{equation}
\label{gauge}
W^a_{(R,H),\mu}\frac{\tau^a}{2}=\frac{1}{2}
\left(\begin{array}{cc}
W^3_{(R,H),\mu} & \sqrt{2}W_{(R,H),\mu}^{+}  \\           
\sqrt{2}W_{(R,H),\mu}^{-} & -W^3_{(R,H),\mu}
\end{array}\right) .
\end{equation}

As mentioned above, the Higgs fields that can accomplish this breaking 
should carry the quantum numbers of $SU(2)_R \otimes SU(2)_H \otimes U(1)_S$. They are $\Delta_{R}= (4,1,3,1)$ and $\Delta_{H} = (4,1,1,3)$. The components of these Higgs fields which can acquire a VEV are the following color-singlet and neutral scalars: $\Delta_{R,H}^{0}$. From (\ref{DelRH2}), one finds:

\begin{equation}
\label{DelVEV}
\langle \Delta_{R,H}^{S} \rangle =
\left(
\begin{array}{cc}
0& \langle \Delta_{R,H}^{0}\rangle/\sqrt{2}=\delta_{R,H}/\sqrt{2}\\
0&0\\
\end{array}
\right) \,.
\end{equation}

The next step is to calculate the masses of the neutral gauge bosons.
This is shown in Appendix A. Here we just quote the results.

The mass matrix squared for the three neutral gauge bosons is found in Appendix A to be
\begin{equation}
\label{massmatrix}
{\cal M}^2_{0}=
\left(\begin{array}{ccc}
g_W^2\delta_R^2 & 0 & -\sqrt{3/8}g_W \tilde g_S\delta_R^2  \\           
0 &g_W^2\delta_H^2& -\sqrt{3/8}g_W \tilde g_S\delta_H^2    \\
\sqrt{3/8} g_W \tilde g_S\delta_R^2&-\sqrt{3/8}g_W\tilde{g}_S\delta_H^2&
({3}/{8})\tilde{g}_S^2(\delta_R^2 +\delta_H^2)
\end{array}\right).
\end{equation}
It is straightforward to diagonalize this matrix. 
The expressions for the general results are rather long and are given
in Appendix A. Here we just present a special case which is quite
reasonable in its own right, namely
\begin{equation}
\label{special}
\delta_R = \delta_H =\delta  \,.
\end{equation}
With (\ref{special}) and Appendix A, we obtain the following mass
eigenstates and eigenvalues.

1) $B_{\mu}$:

\begin{equation}
\label{hyper}
B_{\mu} =\cos \theta_S  \,\tilde{A}_{\mu S} + 
\frac{\sin \theta_S}{\sqrt{2}}\, (W_{\mu R}^3 + W_{\mu H}^3),\quad M_{B}=0,
\end{equation}
where $\cos \theta_S$ and $\sin \theta_S$ are defined in (\ref{thetas}).

2) $Z_{1\mu}$:

\begin{equation}
\label{Z1}
Z_{1\mu} = \sin \theta_S  \,\tilde{A}_{\mu S} -
\frac{\cos \theta_S}{\sqrt{2}}  \,(W_{\mu R}^3 + W_{\mu H}^3), 
\quad M_{Z_{1}} = \frac{1}{2}\,\sqrt{3\, \tilde g_S^2 + 4\,g_W^2}\, \delta.
\end{equation}

3) $Z_{2\mu}$:

\begin{equation}
\label{Z2}
Z_{2\mu} = \frac{1}{\sqrt{2}}\,(W_{\mu H}^3 - W_{\mu R}^3),\quad 
M_{Z_{2}} = g_W \delta.
\end{equation}
 From (\ref{Z1}) and (\ref{Z2}), one obtains also the interesting
relationship in (\ref{formula1}). 

Apart from a slight difference in expressions and notation, the 
above presentation
is very similar to the one given in \cite{HBB} for the group
$SU(4)_{PS} \otimes SU(2)^4$.

\section{Renormalization Group Analysis}
\setcounter{equation}{0}
\subsection{Preliminaries}
In this section we will generalize the renormalization group (RG) analysis of
the gauge couplings and of $\sin^{2}\theta_{W}$
presented in \cite{BH03} to the two-loop level. Moreover we will 
include the contributions of the Higgs scalars to the relevant 
$\beta$-functions that were 
neglected there except for the standard Higgs doublet. At the two-loop level the scalars contribute to the running of the gauge couplings both via gauge as well as Yukawa coupling. The prime goal of this
analysis is to check whether the inclusion of these new effects does not
spoil the early unification of quark and leptons analyzed at one-loop level 
in \cite{BH03}. In calculating the relevant $\beta$ functions we used 
the general formulae of \cite{Machacek}.

We will consider, as in \cite{BH03}, two scenarios. One with $M=\tilde M$ 
and the other with $M>\tilde M$. In both 
scenarios we will set the masses of new fermions in (\ref{NHF}) 
 as well as those of the Higgs fields (\ref{phi}), (\ref{phih}) and
(\ref{phi15}) to be equal to a single scale $M_F$ with 
\be\label{MF}
M_F=M_{\Phi,\Phi_H,\phi^{\beta}}=(250\pm 50)\gev,
\ee
while we will assume that the  scalars in (\ref{delr}) have masses very close to $\tilde M$ so 
that their contributions to the gauge couplings evolution for 
renormalization scales $\mu\le \tilde M$  can be to first approximation 
neglected. On the other hand, these contributions must be taken into account 
for scales $\mu$ in the range  $\tilde M\le \mu\le  M$.

In both scenarios, that is $M=\tilde M$ and $M>\tilde M$, we will use as the
experimental inputs the values of
\be\label{inputq}
\alpha_1(M^2_Z)=\frac{\alpha(M^2_Z)}{\cos^{2}\theta_{W}(M_{Z}^{2})}, 
\qquad
\alpha_2(M^2_Z)=\frac{\alpha(M^2_Z)}{\sin^{2}\theta_{W}(M_{Z}^{2})},
\qquad 
\alpha_3(M_Z^2)
\ee
with 
the $\overline{\rm MS}$ values \cite{PDG}
\begin{equation}\label{newin1}
1/\alpha(M^2_Z)= 127.934(27), \quad 
\alpha_3(M^2_Z)= 0.1172(20),
\end{equation}
\begin{equation}\label{newin2}
\sin^{2}\theta_{W}(M_Z^2)|_{\rm exp}= 0.23113(15)~. 
\end{equation}
The couplings $\alpha_i$ will then be evolved by means of the RG equations
presented below to the scale $M_F$ at which the contributions of the new
heavy fermions (\ref{NHF}) and of the Higgs fields (\ref{phi}), 
(\ref{phih}) and (\ref{phi15}) 
to the $\beta$ functions will be switched on. 
Again, in order to streamline the notation we will denote 
the usual $\alpha^\prime$ 
by $\alpha_1$ and $\alpha_{\rm QCD}$ by $\alpha_3$, reserving 
the index ``$S$" 
for the $U(1)_S$ group.

In the scenario with $M=\tilde M$, the unification scale is simply found from
the petite unification condition (\ref{e2p})
\be\label{con1}
\frac{1}{[\alpha_1(\tilde{M}^{2})]} =\frac{C^2_W}
{[\alpha_2(\tilde{M}^{2})]} +\frac{C_{S}^2}
{[\alpha_{3}(\tilde{M}^{2})]}\, ,
\end{equation}
with $C_W^2$ and $C^2_S$ given in (\ref{CC}). 
In what follows it is useful to denote the l.h.s of (\ref{con1}) 
by $F_{L}$ and the r.h.s. by $F_{R}$, that is
\begin{equation}\label{FLFR}
F_{L} = \frac{1}{[\alpha_1(\tilde{M}^{2})]} \qquad \textrm{and} \qquad F_{R} = 
\frac{C^2_W}{[\alpha_2(\tilde{M}^{2})]} 
+\frac{C_{S}^2}{[\alpha_{3}(\tilde{M}^{2})]}.
\end{equation}

\begin{figure}
\begin{center}
 \epsfig{figure=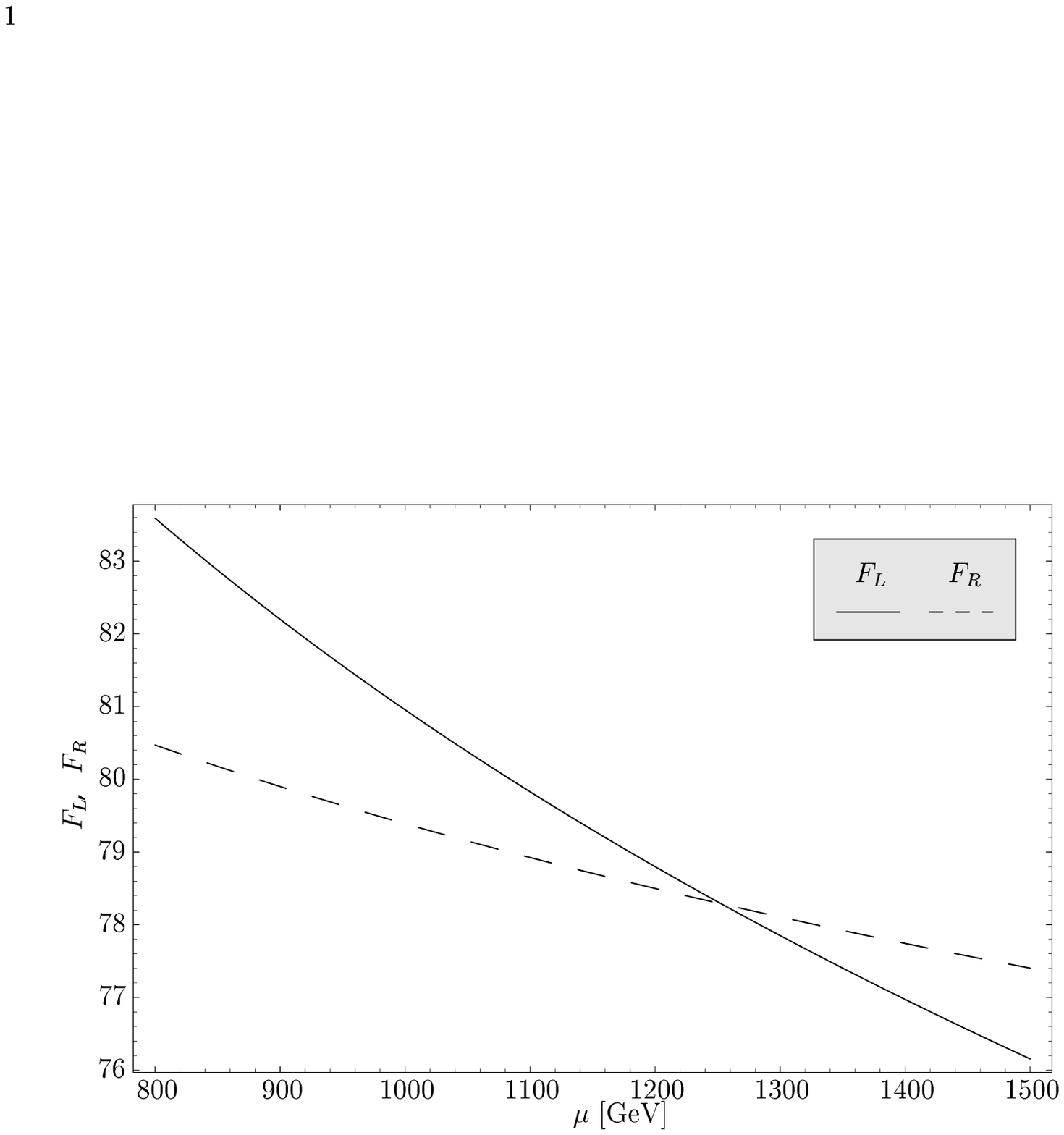,height=8.5cm,angle=0}
 \caption{$F_L$ and $F_R$ in (\ref{FLFR}) as functions of the renormalization
scale $\mu$ for central input variables. 
The unification scale for $M = \tilde{M}$ is uniquely 
determined by the crossing point.}
 \label{diagram1}
\end{center}
\end{figure}

In obtaining (\ref{con1}) we have used $\tilde\alpha_S(\tilde 
M^2)=\alpha_3(\tilde M^2)$ that is valid for $\tilde M=M$. The condition 
(\ref{con1}) corresponds to the equality of the properly normalized gauge
couplings in the standard GUTs as $SU(5)$ and $SO(10)$. Graphically the scale
$\tilde M$ is found as depicted in Figure \ref{diagram1}, 
where we plot $F_L$ and $F_R$ as functions of the scale $\mu$. 
The crossing point of these two evolutions determines uniquely the 
petite unification scale $M=\tilde M$.

\begin{figure}
\begin{center}
 \epsfig{figure=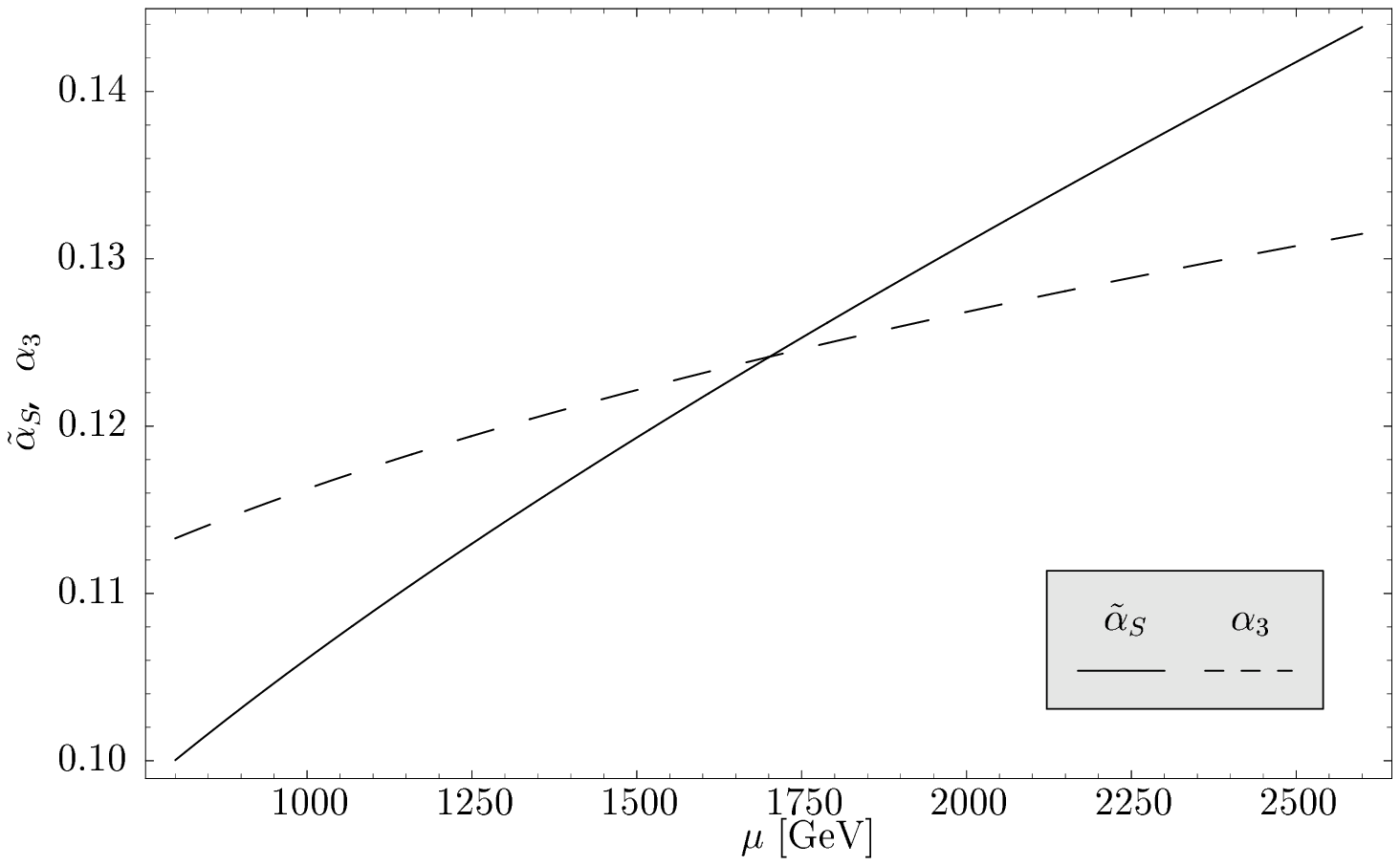,height=8.5cm,angle=0}
 \caption{$\alpha_3$ and $\tilde \alpha_S$ as functions of $\mu>\tilde M$ 
for central input variables.
 The unification scale for $M > \tilde{M}$ is uniquely determined by 
the crossing point.}
 \label{diagram2}
\end{center}
\end{figure}

In the scenario with $M>\tilde M$, the evolution of $\alpha_1$, $\alpha_2$
and $\alpha_3$ for scales $\mu < \tilde M$ is as in the scenario with 
$M=\tilde M$, but now the condition (\ref{con1}) is replaced by    
\be\label{con2}
\frac{1}{[\alpha_1(\tilde{M}^{2})]} =\frac{C^2_W}
{[\alpha_2(\tilde{M}^{2})]} +\frac{C_{S}^2}
{[\tilde\alpha_{S}(\tilde{M}^{2})]}\, ,
\end{equation}
in accordance with (\ref{NHF}),
with $\tilde\alpha_S$ corresponding to the $U(1)_S$ group.
Formula (\ref{con2}) allows to determine the value of 
$\tilde\alpha_S(\tilde M^2)$ at a given scale $\tilde M$ that should be 
larger than $800\gev$ in order to be consistent with the lower bound on the
right--handed gauge boson masses.

Above $\tilde M$ the gauge symmetry group is $SU(3)_c\otimes U(1)_S\otimes
SU(2)^3$  and the evolution of the gauge couplings must also include the
contributions of the Higgs scalars (\ref{delr}). The unification scale
$M$ is then simply found from the condition
\be\label{con3}
\alpha_3(M^2)=\tilde\alpha_S(M^2).
\ee
This is illustrated graphically in Figure \ref{diagram2}. As at the two--loop level the
evolutions of the couplings $\alpha_3$ and $\tilde\alpha_S$ are affected by
the three $SU(2)$ gauge couplings we also need their values at $\tilde M$. 
This is simply found from
\be\label{con4}
\alpha_{2R}(\tilde M^2)=\alpha_{2H}(\tilde M^2)=\alpha_{2L}(\tilde M^2)
\ee
with $\alpha_{2L}(\tilde M^2)\equiv\alpha_2(\tilde M^2)$ calculated as in the scenario with $M=\tilde M$.

In what follows we will first give the RG equations relevant for the case 
$M=\tilde M$. Subsequently we will study the scenario with $M>\tilde M$.

\boldmath
\subsection{Renormalization Group Equations $(M=\tilde M)$} 
\subsubsection{The Range $M_Z\le\mu\le M_F$}
\unboldmath

The relevant RG equations for the evolution of the couplings $\alpha_1$, 
$\alpha_2$ and $\alpha_3$ are given as follows
\be\label{RG1}
\mu\frac{d\alpha_i}{d\mu}=
-\frac{\alpha_i^2}{2\pi}\left[(\beta_0)_i + 
\sum_{j=1,2,3} (\hat\beta_1)_{ij}\frac{\alpha_j}{4\pi}+ 
(\beta_1)^Y_{it}\frac{\lambda_t}{4\pi}\right]
\ee
with $i=1,2,3$ and $\lambda_t=g^2_t/4\pi$, with $g_t$ being the Yukawa
coupling of the top quark. We neglect the contributions from Yukawa couplings
of lighter quarks and from the standard leptons. Moreover as the evolution of
the Yukawa couplings is rather involved at the two-loop level, we will keep
them at constant values corresponding to $m_t(m_t)$ in the case of the top
quark with the same procedure for new heavy quarks and leptons discussed
below. This turns out to be a very good approximation in our case, where the
evolution of couplings takes place over a rather short range of scales, but
of course such a procedure could not be justified in the case of GUTs.

The coefficients $(\beta_0)_i$, $(\hat\beta_1)_{ij}$ and $(\beta_1)^Y_{it}$ are well known but one has to remember that the $U(1)_Y$ coupling $\alpha_1$ used here is differently normalized than the $\alpha_1$ coupling in the $SU(5)$ model. With three generations of quarks and leptons we have then
\be\label{B0}
(\beta_0)_1=-\frac{20}{3}, \qquad  (\beta_0)_2=\frac{10}{3}, \qquad
(\beta_0)_3=7,
\ee

\begin{equation}\label{betaM}
\hat\beta_1=
\left(\begin{array}{ccc}
-95/9&-3& -44/3\\
-1& -11/3   & -12\\
-11/6 & -9/2 & 26
\end{array}\right)
\end{equation}
and
\be\label{BY0}
(\beta_1)^Y_{1t}=\frac{17}{6}, \qquad  (\beta_1)^Y_{2t}=\frac{3}{2}, \qquad
(\beta_1)_{3t}^Y=2~.
\ee

\boldmath
\subsubsection{The Range $M_F\le\mu\le \tilde M$}
\unboldmath
Above the scale $M_F$ the contributions of new fermions in (\ref{NHF})
 and the scalars in (\ref{phi}), (\ref{phih}) and 
(\ref{phi15}) have to be taken into account. This modifies the coefficients 
in (\ref{B0}) and (\ref{betaM}) as follows
\be\label{B0F}
(\beta_0)_1=-\frac{136}{3}-\frac{209}{9}, \qquad  (\beta_0)_2=-\frac{2}{3}-\frac{19}{3}, \qquad
(\beta_0)_3=3-\frac{16}{3},
\ee
\begin{equation}\label{betaMF}
\hat\beta_1=
\left(\begin{array}{ccc}
-2230/9-17777/27&-42-209& -280/3-3200/9\\
-14-209/3& -158/3-247/3   & -24-128\\
-35/3-400/9 & -9-48 & -50-592/3
\end{array}\right),
\end{equation}
where second terms in (\ref{B0F}) and (\ref{betaMF}) represent scalar contributions. In addition Yukawa couplings of new quarks and leptons have to be 
taken into account. With respect to the democratic fermion mass model 
presented in Section \ref{yukawacouplings} the Yukawa term in the RG 
equations (\ref{RG1}) is generalized to
\be\label{YUK}
\Delta_Y\left(\mu\frac{d\alpha_i}{d\mu}\right)=
-\frac{\alpha_i^2}{2\pi}\sum_k
(\beta_1)^Y_{ik}\frac{\lambda_k}{4\pi},
\ee
where the sum runs over 13 Yukawa couplings corresponding to seven heavy
quarks and six heavy leptons. 
The coefficients $(\beta_1)^Y_{it}$ are as in (\ref{BY0}).
The coefficients $(\beta_1)^Y_{ik}$ with $k \neq t$
relevant for the contributions of new
fermions  are found to be
\begin{itemize}
\item with respect to $SU(3)_c$:
\be
(\beta_1)^Y_{3\tilde Q}=2, \qquad  
\tilde Q=\tilde U,\tilde D,\tilde S,\tilde C,\tilde B,\tilde T
\ee
\item with respect to $SU(2)_L$:
\begin{eqnarray}
(\beta_1)^Y_{2\tilde Q} & = &\frac{3}{2}, \qquad  
\tilde Q=\tilde U, \tilde D,\tilde S,\tilde C,\tilde B,\tilde T \nonumber \\
(\beta_1)^Y_{2\tilde L}& = &\frac{1}{2}, \qquad  
\tilde L=\tilde l_u,\tilde l_d, \tilde l_s,\tilde l_c, \tilde l_b, 
\tilde l_t
\end{eqnarray}

\item with respect to $U(1)_Y$:
\begin{eqnarray}
(\beta_1)^Y_{1\tilde Q} & = &\frac{89}{6}, \qquad  \tilde Q=\tilde U,\tilde
C,\tilde T \nonumber \\
(\beta_1)^Y_{1\tilde Q}& = &\frac{29}{6}, \qquad  \tilde Q=\tilde D,\tilde
S,\tilde B \nonumber \\
(\beta_1)^Y_{1\tilde L} & = &\frac{13}{2}, \qquad  
\tilde L=\tilde l_u,\tilde l_c, \tilde l_t \nonumber \\
(\beta_1)^Y_{1\tilde L} & = &\frac{25}{2}, \qquad  
\tilde L=\tilde l_d, \tilde l_s, \tilde l_b~. 
\end{eqnarray}
\end{itemize}

It is clear from these formulae that the Yukawa couplings of 
$(\tilde U,\tilde C,\tilde T)$, $(\tilde D,\tilde S,\tilde B)$, 
$(\tilde l_u,\tilde l_c, \tilde l_t)$ and 
$(\tilde l_d, \tilde l_s, \tilde l_b)$, 
generally differ from each other. However, in order to get a rough estimate
of Yukawa contributions let us set all Yukawa couplings of the new fermions
to be equal to the top Yukawa coupling $g_t$. Then effectively only the 
$\lambda_t$ term is present as in (\ref{RG1}) but the coefficients 
$(\beta_1)^Y_{it}$ are modified as follows
\be\label{BY0F}
(\beta_1)^Y_{1t}=\frac{713}{6}, \qquad  (\beta_1)^Y_{2t}=\frac{27}{2}, \qquad 
(\beta_1)_{3t}^Y=14.
\ee
\boldmath
\subsubsection{Analytical Formula for $M$}
\unboldmath
We will present the numerical analysis of this scenario in subsection 4.4. On
the other hand it is possible to derive an approximate analytical formula for
$M$ as function of the input parameters (\ref{MF}), (\ref{newin1}) and 
(\ref{newin2}). To this end we define effective ``one loop" coefficients 
\be\label{eff1}
\beta_i^{\rm eff}\equiv (\beta_0)_i + 
\sum_{j=1,2,3} (\hat\beta_1)_{ij}\frac{\alpha_j}{4\pi}
+(\beta_1)^Y_{it}\frac{\lambda_t}{4\pi},
\ee
with $\alpha_j$ and $\lambda_t$ frozen at some intermediate scale
$M_Z\le\mu\le M_F$. 
Analogous procedure is used for the range $M_F\le \mu \le \tilde M=M$.

We then find
\be\label{master1}
M=M_F\left[\frac{M_Z}{M_F}\right]^{K_3/K_{\rm tot}} \exp(P),
\ee
where
\be\label{Pdef}
P=\frac{1}{8\pi\alpha(M^2_Z) K_{\rm tot}}
\left(1-C^2_S\frac{\alpha(M^2_Z)}{\alpha_3(M^2_Z)} -
\frac{\sin^{2}\theta_{W}(M_Z^2)}{\sin^{2}\theta_{W}^0}\right)
\ee
and
\be\label{def-K3}
K_3=\frac{1}{16\pi^2}\left(\beta_3^{\rm eff}C_S^2+\beta_2^{\rm eff}C_W^2
    -\beta_1^{\rm eff}\right)
\ee
with the RG coefficients given in (\ref{B0})--(\ref{betaM}). 
Formula (\ref{def-K3}) is also valid for $K_{\rm tot}$ but this time 
the coefficients (\ref{B0F})-(\ref{BY0F}) relevant for 
the range $M_F\le\mu\le \tilde M$ should
be used. The formula (\ref{master1}) can be directly obtained from 
(\ref{sinsq2}) by setting $M=\tilde M$ and 
making the replacement:
\be\label{LOGS}
K\ln\frac{\tilde{M}}{M_Z}\to
K_3\ln\frac{M_F}{M_Z}+ K_{\rm tot}\ln\frac{{M}}{M_F}. 
\ee

\boldmath
\subsection{Renormalization Group Equations ($M>\tilde M)$}
\unboldmath
\subsubsection{Preliminaries}
\begin{figure}
\begin{center}
 \epsfig{figure=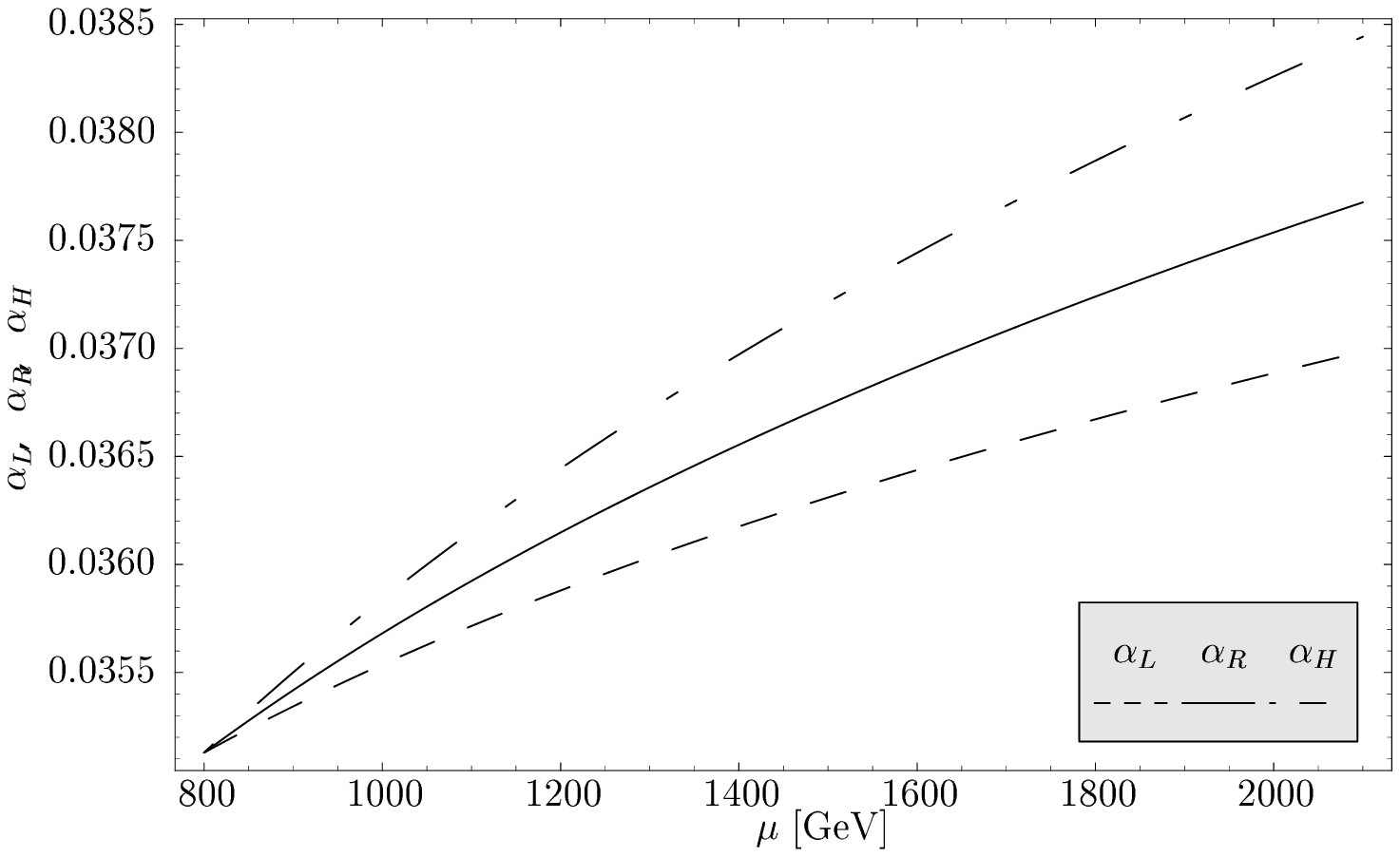,height=8.5cm,angle=0}
 \caption{Running of $\alpha_{L}$, $\alpha_{R}$, $\alpha_{H}$ 
for $M > \tilde{M}$, with all input variables taken at their central values.}
 \label{diagram3}
\end{center}
\end{figure}

The evolution of couplings from $\mu=M_Z$ to $\mu=\tilde M$ proceeds as in
the previous scenario and consequently the formulae given in the subsections
4.1 and 4.2 allow the determination of the couplings $\alpha_3(\tilde M^2)$,
$\tilde\alpha_S(\tilde M^2)$, $\alpha_{2L}(\tilde M^2)$,
$\alpha_{2R}(\tilde M^2)$ and $\alpha_{2 H}(\tilde M^2)$, that constitute 
the starting point for the subsequent evolution from $\tilde M$ to $M$. We already mentioned that the Higgs fields break the equality of the three $SU(2)$ couplings and consequently five different couplings have to be considered, although the splitting of the three $SU(2)$ couplings is insignificant as one can see in Figure \ref{diagram3}.
\boldmath
\subsubsection{The Range $\tilde M\le\mu\le M$}
\unboldmath
The relevant RG equations for the evolution of the couplings 
$\alpha_3(\mu^2)$,
$\tilde\alpha_S(\mu^2)$, $\alpha_{2L}(\mu^2)$,
$\alpha_{2R}(\mu^2)$ and $\alpha_{2 H}(\mu^2)$ are given 
as follows $(i=3,S,2L,2R,2H)$
\be\label{RG2}
\mu\frac{d\alpha_i}{d\mu}=
\left(\mu\frac{d\alpha_i}{d\mu}\right)_G+
\Delta_Y\left(\mu\frac{d\alpha_i}{d\mu}\right),
\ee
where the first term on the r.h.s describes the evolution of the gauge
couplings with all Yukawa couplings set to zero and the second term takes
into account the presence of these couplings.
We then have
\be\label{RG3}
\left(\mu\frac{d\alpha_i}{d\mu}\right)_G=
-\frac{\alpha_i^2}{2\pi}\left[(\beta_0)_i + 
\sum_{j} (\hat\beta_1)_{ij}\frac{\alpha_j}{4\pi}\right],
\ee
with $j$ running over the five couplings and
\be\label{RG4}
\Delta_Y\left(\mu\frac{d\alpha_i}{d\mu}\right)=
-\frac{\alpha_i^2}{2\pi} 
\sum_{r} (\beta_1)^Y_{ir}\frac{\lambda_r}{4\pi},
\ee
as in (\ref{YUK}) but this time i=3, S, 2L, 2R, 2H. 

With three generations of ordinary and heavy quarks and leptons 
the coefficients in
(\ref{RG3}) are given as follows (second terms below stand for scalar contributions):
\begin{eqnarray}\label{B0FF}
(\beta_0)_3=3-\frac{19}{3}  , \qquad  (\beta_0)_S=-8-\frac{19}{3}, \qquad 
(\beta_0)_{2L}=-\frac{2}{3}-\frac{19}{3}, \nonumber \\ 
(\beta_0)_{2R}=-\frac{2}{3}-\frac{27}{3},  \qquad  (\beta_0)_{2H}=-\frac{26}{3}-\frac{16}{3}  
\end{eqnarray}
and with the ordering $i=3,S,2L,2R,2H$
\begin{equation}\label{betaMF1}
\hat\beta_1 = \left(\hat{\beta}_{1}\right)_{GF} + \left(\hat{\beta}_{1}\right)_{H},
\end{equation}
where
\begin{equation}
\left(\hat{\beta}_{1}\right)_{GF} = 
\left(\begin{array}{ccccc}
-50 & -1 & -9 & -9 & -18 \\
-8 & -7 & -9 & -9 & -18 \\
-24 & -3 & -158/3 & 0 & -18 \\
-24 & -3 & 0 & -158/3 & -18 \\
-48 & -6 & -18 & -18 & -452/3
\end{array}\right)
\end{equation}
are the contributions arising from fermions and gauge bosons and
\begin{equation}
\left(\hat{\beta}_{1}\right)_{H} = 
\left(\begin{array}{ccccc}
-658/3 & -67/6 & -48 & -60 & -12 \\
-268/3 & -277/6 & -48 & -60 & -12 \\
-128 & -16 & -247/3 & -57 & -24 \\
-160 & -20 & -57 & -157 & -24 \\
-32 & -4 & -24 & -24 & -448/3
\end{array}\right)
\end{equation}
are contributions coming from the scalars. With the separation of the scalar contributions the  breaking of the symmetry between the $SU(2)$ couplings becomes obvious: $\alpha_{2H}$,$\alpha_{2L}$ and $\alpha_{2R}$ evolve slightly differently. For the coefficients 
$(\beta_1)^Y_{ir}$ we find
\begin{itemize}
\item with respect to $SU(3)_c$:
\be
(\beta_1)^Y_{3Q}=2, \qquad  Q=~{\rm all~quarks}
\ee
\item with respect to $SU(2)_L$:
\begin{eqnarray}
(\beta_1)^Y_{2L\tilde Q} & = &\frac{3}{2}, \qquad  
\tilde Q=t,\tilde U,\tilde D,\tilde S,\tilde C,\tilde B,\tilde T \nonumber \\
%
%
(\beta_1)^Y_{2L\tilde L} & = &\frac{1}{2}, \qquad  
\tilde L=\tilde l_u,\tilde l_d, \tilde l_s,\tilde l_c, \tilde l_b, 
\tilde l_t 
%
\end{eqnarray}
\item with respect to $SU(2)_R$:
\begin{eqnarray}
(\beta_1)^Y_{2R\tilde Q} & = & (\beta_1)^Y_{2L\tilde Q}, \qquad
(\beta_1)^Y_{2R\tilde L}=(\beta_1)^Y_{2L\tilde L}, 
%
\end{eqnarray}
\item with respect to $SU(2)_H$:
\begin{eqnarray}
(\beta_1)^Y_{2H\tilde Q} & = &3, \qquad  
\tilde Q=t,\tilde U,\tilde D,\tilde S,\tilde C,\tilde B,\tilde T \nonumber \\
(\beta_1)^Y_{2H\tilde L} & = & {1}, \qquad  
\tilde L=\tilde l_u,\tilde l_d, \tilde l_s,\tilde l_c, \tilde l_b, 
\tilde l_t
\end{eqnarray}
\item with respect to $U(1)_S$:
\begin{eqnarray}
(\beta_1)^Y_{SQ}=\frac{1}{2}, \qquad  {\rm Q=~all~quarks} \nonumber \\
(\beta_1)^Y_{SL}=\frac{3}{2}, \qquad  {\rm L=~all~leptons}
\end{eqnarray}
\end{itemize}
with all remaining coefficients being zero.

Again as in the case of the range $M_F\le\mu\le\tilde M$ we can get a rough
estimate of the effects of the Yukawa couplings by setting them equal to each
other. In this case
\be\label{RG5}
\Delta_Y\left(\mu\frac{d\alpha_i}{d\mu}\right)=
-\frac{\alpha_i^2}{2\pi} 
(\beta_1)^Y_{i}\frac{\lambda_Y}{4\pi},
\ee
where $\lambda_Y$ is universal and 
\be
(\beta_1)^Y_3=14, \qquad  (\beta_1)^Y_{1S}=\frac{25}{2},
\ee
\be
(\beta_1)^Y_{2L}=(\beta_1)^Y_{2R}=\frac{27}{2},
 \qquad  (\beta_1)^Y_{2H}=27.
\ee
\boldmath
\subsubsection{Analytic Formula for $M$}
\unboldmath
Proceeding as in subsection 4.2 one can derive an approximate analytic
formula for $M$ as a function of $\tilde M$, $M_F$ and the input parameters
(\ref{MF}), (\ref{newin1}) and 
(\ref{newin2}). We find
\be\label{master2}
M=\tilde M\left[\frac{M_Z}{M_F}\right]^{K_3/K^\prime} 
\left[\frac{M_F}{\tilde M}\right]^{K_{\rm tot}/K^\prime}
\exp(P^\prime),
\ee
where
\be\label{Pdefprim}
P^\prime=P\frac{K_{\rm tot}}{K^\prime}
\ee
with
$K_3$, $K_{\rm tot}$ and $P$ defined in subsection 4.2 and
\be\label{def-Kprim}
K^\prime=\frac{C^2_S}{16\pi^2}\left(\beta_3^{\rm eff} -\beta_S^{\rm
eff}\right) 
\ee
where $\beta_3^{\rm eff}$ and $\beta_S^{\rm eff}$ are evaluated as in 
(\ref{eff1}) but with various coefficients relevant for the range 
$\tilde M\le\mu\le M$. 
\subsection{Numerical Analysis}
We have solved the RG equations listed above numerically to find $M=\tilde M$
in the first scenario and $M$ in the scenario with $M > \tilde M$ and 
$\tilde M \ge 800\gev$. Let us first neglect the Yukawa contributions.
The dependence of $M$ on the input parameters is in 
the first case as follows: 
\begin{itemize}
\item
As $\sin^{2}\theta_{W}(M_{Z}^{2})$ is very precisely known, the variation of
its value within the full range in (\ref{newin2}) introduces only a  shift 
in the ballpark of $30\gev$. $M$ decreases with increasing 
$\sin^{2}\theta_{W}(M_Z^2)$.
\item
The uncertainty of $M$ due to $M_F$ and $\alpha_3(M^2_Z)$ is significant 
as shown in table~\ref{tab1}, where we have set $\alpha(M^2_Z)$ and
$\sin^{2}\theta_{W}(M_Z^2)$ at their central values and varied $M_F$  
and $\alpha_3(M^2_Z)$ in the ranges given in (\ref{MF}) and 
(\ref{newin1}). $M$ increases with increasing $M_F$ and $\alpha_3(M^2_Z)$.
\item
The effect of the NLO contributions is significant. They increase the
scale $M=\tilde M$ by roughly 160$\gev$ 
\end{itemize}

In the case $M>\tilde M$ our findings are as follows:
\begin{itemize}
\item
As expected, the value of $M$ increases with 
decreasing $\tilde M$. In table~\ref{tab1} the last two columns show the 
LO and NLO values of $M$ for $\tilde M=800\gev$. The pattern of the 
dependence on $M_F$  and $\alpha_3(M^2_Z)$ is similar to the one in the case 
$M=\tilde M$.
\item
The NLO corrections in this case are even more important and amount to the
increase of $M$ by $370\pm 40\gev$ for $\tilde{M}=800\gev$ with 
a smaller shift for $\tilde{M}>800\gev$.
\end{itemize}

Most importantly
\begin{itemize}
\item
The inspection of the renormalization group coefficients of Sections 4.2 and
4.3 shows that the large impact of NLO corrections originates to a large
extent in the scalar contributions.
\item
Moreover, as discussed below, the latter contributions increase $M$  
roughly by $250\gev$ and $450\gev$ for $M=\tilde M$ and 
$M>\tilde{M}=800\gev$, respectively.
\end{itemize}

\begin{table}[hbt]
\vspace{0.4cm}
\begin{center}
\caption[]{\small \label{tab1} Values for the unification scale $M$
for various values of $\alpha_3(M^2_Z)$ and $M_F$, 
$1/\alpha(M^2_Z)= 127.934$  and 
$\sin^{2}\theta_{W}(M_Z^2)= 0.23113)$. In the last two columns 
we set $\tilde M=800\gev$. The RGEs have been set up including gauge, 
fermionic and scalar contributions.
}
\begin{tabular} {|c|c||c|c||c|c|}\hline
 $\alpha_3(M^2_Z)$ & $M_F$ & $M=\tilde M,~LO$ & $M=\tilde M,~NLO$ 
 & $M> \tilde M,~LO$ & $M>\tilde M,~NLO$  \\ \hline\hline 
        & 200 & 961.4 & 1120.9 & 1083.0 & 1410.4 \\ 
 0.1152 & 250 & 1036.7 & 1180.8 & 1226.2 & 1538.6 \\ 
        & 300 & 1102.6 & 1233.7 & 1357.2 & 1655.6 \\ \hline 
        & 200 & 1012.1 & 1190.4 & 1178.5 & 1562.3 \\ 
 0.1172 & 250 & 1091.3 & 1253.0 & 1334.5 & 1701.6 \\ 
        & 300 & 1160.7 & 1308.2 & 1477.0 & 1828.9 \\ \hline 
        & 200 & 1063.6 & 1262.3 & 1279.0 & 1726.4 \\ 
 0.1192 & 250 & 1146.9 & 1327.3 & 1448.2 & 1877.3 \\  
        & 300 & 1219.7 & 1385.0 &  1602.9 & 2015.2 \\ \hline 
\end{tabular}
\end{center}
\end{table}

In summary, varying all the input parameters within one standard deviation 
we find at the NLO level
\be\label{RES1}
M=(1253\pm 132)\gev, \qquad (M=\tilde M)
\ee
and
\be\label{RES2}
M=(1713\pm 302)\gev, \qquad (M>\tilde M = 800\gev).
\ee

\subsection{Anatomy of Renormalization Group Effects }
In what follows we would like to present the results of a number of exercises
that we made in the context of our numerical analysis.

First the inclusion of Yukawa couplings changes the values in
(\ref{RES1}) and (\ref{RES2}) respectively to
\be\label{RES1a}
M=(1273\pm 132)\gev, \qquad (M=\tilde M)
\ee
and
\be\label{RES2a}
M=(1742\pm 300)\gev, \qquad (M>\tilde M= 800\gev).
\ee
The relatively small impact of Yukawa couplings on our results justifies our
rough treatment of these contributions.

Next, removing the scalar contributions altogether we find instead of
(\ref{RES1}) and (\ref{RES2})
\be\label{RES1b}
M=(994\pm 120)\gev, \qquad (M=\tilde M)
\ee
and
\be\label{RES2b}
M=(1254\pm 300)\gev, \qquad (M>\tilde M= 800\gev),
\ee
implying that these contributions are very important.

Finally, we have investigated the impact of the increase of $M_F$ to
$M=\tilde M$. This is equivalent to the removal of the heavy fermion
contributions to the renormalization group equations below the unification
scale. We find then for the maximal unification scale obtained with
largest $\alpha_3(M_Z^2)$ and smallest $\sin^2\theta_W(M_Z^2)$
\be
M_{\rm max}=\tilde M= M_F = 4595~ (2442)\gev
\ee
with and (without) scalar contributions (at the scale 250\gev), respectively. Thus even without heavy
fermions below $\mu=M$ we obtain a relatively low unification scale.
That is, the group theoretic factors $C_S$ and $C_W$ in our master formula 
guarantee early unification even if only the SM fields have masses below 
the unification scale. On the other hand the presence of
new heavy fermions with masses $\mathcal{O}\left(250 \, \rm{GeV} \right)$ is necessary in order to keep $M$ below $2~\rm{TeV}$.



\section{The Fate of the Unconventional Fermions}
\label{unconventional}
\setcounter{equation}{0}
\subsection{Relevant Interactions}

As we have shown in \cite{BH03} and in Section 2.4, the construction
of the model necessitates the introduction of new heavy quarks and leptons with
unconventional charges: $\pm 4/3$ for the quarks and $\pm 2$ for the leptons.
These fermions behave in exactly the same manner as the ordinary fermions
and are linked to the latter by $SU(2)_H$ gauge interactions
$g_{W} \bar{\Psi}_{L,R} \vec{W}_{\mu H} . \vec{T}_H\,\gamma^{\mu} \Psi_{L,R}$,
and by the Yukawa interactions $g_{\Phi_H} \bar{\Psi}_L \vec{\Phi}_{H} . \vec{T}_H 
\Psi_R$. The question arises as to the fate of the lightest of these new fermions.

One should note that the vector-like quarks and leptons as written down in 
\cite{BH03} and (\ref{VF}) suffer from problems with cosmological constraints in the 
following way. Since they carry electric charges such as 5/6 or 3/2, the
lightest one would be absolutely stable since it cannot decay into 
known particles. There are severe constraints on objects such as stable 
fractionally charged leptons \cite{goldberg} and, unless the 
numbers are severely
depleted by exotic mechanisms such as the one proposed in \cite{goldberg},
they are ruled out observationally. We will therefore omit them altogether.

Let us first discuss the fate of the unconventional
$SU(2)_H$ partners of standard quarks and leptons 
by looking only at the gauge interactions.
For the lightest new quark, if the gauge bosons
{\em do not} mix, one would face the dreadful conclusion that it would be
absolutely {\em stable}. We will discuss below various constraints which rule
out that situation. Fortunately, we will see that there are Higgs fields
whose VEVs mix the gauge bosons of the different groups and it is this mixing
which renders the lightest new quark unstable, i.e. it can now decay into
conventional fermions which are assumed to be lighter. A similar
argument applies to the case of the new leptons.

In this section, we focus only on the charge-changing interactions
and hence on the charged gauge bosons of $SU(2)_{L} \otimes
SU(2)_{R} \otimes SU(2)_{H}$. We have already discussed above the breaking
of $SU(2)_{R} \otimes SU(2)_{H}$ and, in particular, the neutral gauge
boson sector. In that discussion, we have used the Higgs fields
$\Delta_R=(4,1,3,1)$ and $\Delta_H=(4,1,1,3)$. However, as we have 
shown in Sections \ref{choice} and \ref{charges}, other Higgs fields
come into play, namely $\Phi= (1,2,2,1)$, $\Phi_H = (1,2,2,3)$ as
well as $\phi^{(1)}$ which is the color singlet part of
$\phi^{\beta}= (15,2,2,1)$. The VEVs of these scalars will evidently
mix various gauge bosons. We have also mentioned earlier that, for
symmetry reasons, one would also like to have $\Delta_L = (4,3,1,1)$,
$\tilde{\Phi}_1 = (1,2,1,2)$ and $\tilde{\Phi}_2 = (1,1,2,2)$. A detailed
treatment of this problem is given in Appendix A. Here we will give
a simplified version for illustration. In particular, we will
make the same assumption as in (\ref{special}), namely
$\delta_R = \delta_H =\delta $. Furthermore, we will assume that
$\delta > \langle \tilde{\Phi}_2 \rangle \gg 
\langle \Phi \rangle, \,\langle \Phi_H \rangle, \,
\langle \tilde{\Phi}_1 \rangle, \, \langle \phi^{(1)} \rangle$.
For simplicity, we will also assume that 
$\langle \Delta_L \rangle = 0$. Below, we will denote a generic
electroweak scale by $v$ and a generic large scale by $\delta$.

Since this section focuses primarily on the question of whether or not
the lightest of the unconventional fermions are stable, we will present
a streamlined version of the discussion which will accurately summarize
the points that we wish to make. Some of the details can be found in 
Appendix A.

The lightest of the unconventional fermions can only decay
into either a conventional fermion plus the SM $W$ boson or
into three conventional fermions if it has the appropriate mass.
This implies that the current $J^{\mu}_{H}$ should interact
with the SM $W$ for the former case or it mixes with either
$J^{\mu}_{L}$ or $J^{\mu}_{R}$ for the latter case. This is what
we will show below. We use the notation $g_W$ for the
three gauge couplings which are defined to be equal to each other
at $\tilde{M}$ (see (\ref{g2}). In the estimate of the decay
rates given below, we shall, however, take the value of $g_2$ or equivalently
of the Fermi constant $G_{F}$ for simplicity.

With the above remarks taken into account, we now list the eigenvalues
and mass eigenstates for the charged gauge bosons. (The exact expressions
with various factors included are given in Appendix A.) Our notations
are as follows. We use $\tilde{W}_L, W_1, W_2$ for the mass eigenstates
and $W_L, W_R, W_H$ for the gauge eigenstates. We have
\be
\label{massl}
M_{\tilde{W}_L} = \mathcal{O}(g_W\,v),
\quad M_{W_1} = \mathcal{O}(g_W\,\delta),\quad M_{W_2} =\mathcal{O}(g_W\,\delta).
\ee
The relationship between the gauge and mass eigenstates is given by
\begin{equation}
\label{Weigen}
\left(\begin{array}{c}
W_L^{\pm} \\ W_R^{\pm} \\ W_H^{\pm}
\end{array}\right)= 
O^T\left( \begin{array}{c}
\tilde{W}_{L}^{\pm} \\W_{1}^{\pm} \\ W_{2}^{\pm}
\end{array} \right) =
\left(\begin{array}{ccc}
1&{\cal{O}}(v^2/\delta^2)&{\cal{O}}(v^2/\delta^2) \\
{\cal{O}}(v^2/\delta^2)&\frac{1}{\sqrt{2}}& \frac{1}{\sqrt{2}}\\
{\cal{O}}(v^2/\delta^2)&\frac{1}{\sqrt{2}}&-\frac{1}{\sqrt{2}}
\end{array} \right)
\left( \begin{array}{c}
\tilde{W}_{L}^{\pm} \\W_{1}^{\pm} \\ W_{2}^{\pm}
\end{array} \right).
\end{equation}

Let us denote generically the charge-changing currents associated with
$SU(2)_L$, $SU(2)_R$, and $SU(2)_H$ by $J^{\mu}_{L}$, $J^{\mu}_{R}$,
and $J^{\mu}_{H}$ respectively.
The charged current interactions
involving these currents can now be written as (with $\pm$ omitted for
simplicity)
\be
\label{ccinter1}
{\cal L}_{interaction} = g_W (J^{\mu}_{L}, J^{\mu}_{R},
J^{\mu}_{H})\times \left(\begin{array}{c}
W_{\mu\,L} \\ W_{\mu\,R} \\ W_{\mu\,H}
\end{array}\right) \,.
\ee
In terms of the gauge boson mass eigenstates, one can now write (\ref{ccinter1})
as
\be
\label{ccinter2}
{\cal L}_{interaction} = g_W (\tilde{J}^{\mu}_{L}, \tilde{J}^{\mu}_{R}, 
\tilde{J}^{\mu}_{H}) \times \left(\begin{array}{c}
\tilde{W}_{\mu\,L} \\ W_{\mu\,1} \\ W_{\mu\,2}
\end{array}\right) \,,
\ee
where
\be
\label{cuurent}
(\tilde{J}^{\mu}_{L}, \tilde{J}^{\mu}_{R}, \tilde{J}^{\mu}_{H})=
(J^{\mu}_{L}, J^{\mu}_{R}, J^{\mu}_{H}) \times O^T \,,
\ee
and where the rotation matrix $O^T$ is given in (\ref{Weigen}). Explicitely,
we now list separately the following interactions.

\bi

\item Interaction with $\tilde{W}_{\mu\,L}$:
\begin{eqnarray}
\label{WtilintL}
{\cal L}_L& =& g_W\,\tilde{J}^{\mu}_{L}\tilde{W}_{\mu\,L} \nonumber \\
         &=& g_W[J^{\mu}_{L} + {\cal{O}}(v^2/\delta^2)(J^{\mu}_{R}+
J^{\mu}_{H})]\tilde{W}_{\mu\,L}
\end{eqnarray}

\item Interaction with $W_{\mu\,1}$:
\begin{eqnarray}
\label{Wint1}
{\cal L}_1& =& g_W\,\tilde{J}^{\mu}_{R} W_{\mu\,1} \nonumber \\
         &=& g_W[{\cal{O}}(v^2/\delta^2)J^{\mu}_{L} + \frac{1}{\sqrt{2}}
(J^{\mu}_{R}+ J^{\mu}_{H})]W_{\mu\,1}
\end{eqnarray}

\item Interaction with $W_{\mu\,2}$:
\begin{eqnarray}
\label{Wint2}
{\cal L}_2& =& g_W\,\tilde{J}^{\mu}_{H} W_{\mu\,2} \nonumber \\
         &=& g_W[{\cal{O}}(v^2/\delta^2)J^{\mu}_{L} + \frac{1}{\sqrt{2}}
(J^{\mu}_{R}- J^{\mu}_{H})]W_{\mu\,2}
\end{eqnarray}

\ei

A few remarks are in order here.
\bi

\item From (\ref{WtilintL}), we observe that the contribution of
a $V+A$ current to known weak interactions is suppressed in the
interaction Lagrangian by a factor ${\cal{O}}(v^2/\delta^2)$.
Experimental constraints \cite{PDG} from searches for right-handed
currents in normal weak interactions give ${\cal{O}}(v^2/\delta^2)
< 10^{-3}$. This can easily be satisfied within our model through
the choices of the various VEVs.

\item Also from (\ref{WtilintL}), one can see that $J^{\mu}_{H}$, which
changes a conventional fermion into an unconventional one and vice versa,
couples with $\tilde{W}_{\mu\,L}$ with a factor ${\cal{O}}(v^2/\delta^2)$.
This has some important implications concerning the decay of the
lightest of the unconventional fermions.

1) If the lightest unconventional fermion which appears in
$J^{\mu}_{H}$ has a mass greater than the sum of the accompanying
conventional fermion mass and the W-boson mass, it can have the
decay mode $\tilde{F} \rightarrow f + W$ ($W$ being the standard
$\tilde{W}_{\mu\,L}$) via the interaction
\be
\label{realW}
g_W{\cal{O}}(v^2/\delta^2)J^{\mu}_{H}\tilde{W}_{\mu\,L}.
\ee

Even though
${\cal{O}}(v^2/\delta^2) < 10^{-3}$, the decay rate can be substantial
because the unconventional fermion decays into a real $W_L$.

2) If the mass of the lightest unconventional fermion is less than
the sum of the $W$-mass and the mass of the accompanying conventional fermion,
the decay can occur through the interaction:
\be
\label{virtualW}
2g_W^2 {\cal{O}}(v^2/\delta^2) J^{\mu}_{L}\, \frac{1}{q^2 - m_{\tilde{W}_L}^2} \, 
J_{\mu H}.
\ee

\item If the mass of the unconventional fermion is between
the sum of the accompanying conventional fermion mass and the $W$-boson mass
and either that of $W_1$ or $W_2$, the decay would be into
a real $W_L$ as in (1) above.

\item  If the mass of the lightest unconventional fermion is less than
the sum of the $W$-mass and the mass of the accompanying conventional fermion 
then, from (\ref{Wint1}) and (\ref{Wint2}), 
one obtains the following additional contributions to the decay:
\be
\label{virtualW12}
g_W^2 J^{\mu}_{R} (\frac{1}{q^2 - m_{W1}^2} - \frac{1}{q^2 - m_{W2}^2})
J_{\mu H}.
\ee

\ei

In computing the matrix elements for various decays, one has to express the 
fermions
which appear in $J^{\mu}_{L}$, $J^{\mu}_{R}$ and $J^{\mu}_{H}$ in terms
of the mass eigenstates. This will result in the appearance of a number of
mixing angles which are different from the CKM elements. In the absence of
a plausible model of fermion masses (especially for the unconventional ones),
the best one can do is to make estimates of the decay rates based on
reasonable assumptions about the magnitudes of these unknown mixing angles.

\subsection{Decay Modes}

We will present here some estimates of possible decay modes
of the {\em lightest} unconventional quark and lepton. The main
purpose, as we have mentioned above, is to see how fast or how slow these
fermions decay. Since there are strong constraints on ``stable'' fermions
from cosmology and from collider experiments, the unconventional fermions
should be sufficiently heavy and should decay fast enough to evade these
bounds. This is what we will show below. A comprehensive study
of all possible decays is beyond the scope of this paper and it
will be included in a future publication. Here we wish to
merely present some illustrative examples.

To be more specific, let us assume that the lightest unconventional 
quark is $\tilde{U}(4/3)$ and the lightest unconventional lepton is
$\tilde{l}_{u}(-1)$, where the numbers inside the parentheses represent
the charges of these particles. (One can easily change this scenario
to, e.g., $\tilde{D}(1/3)$ and $\tilde{l}_{d}(-2)$, or any other
combination.) Since those lightest unconventional fermions cannot
decay into other unconventional fermions, the only option left is
for them to decay into conventional fermions.

In what follows, we will use the following notations for the
conventional fermions: $\psi^{i}_{d}= d,s,b$ and $\psi^{i}_{\nu}
=\nu_e, \nu_{\mu}, \nu_{\tau}$, for
the quarks and the neutral leptons respectively. Notice that
$SU(2)_H$ interactions connect $\tilde{U}(4/3)$ to 
$\bar{\psi}^{i}_{d}$, and $\tilde{l}_{u}(-1)$ to $\psi^{i}_{\nu}$.

In using $J^{\mu}_{H}$, we will express the fields which appear
in that current in terms of their mass eigenstates. Consequently,
the results presented below will contain mixing angles of the type
$V_{\tilde{U}\,\psi^{i}_{d}}$.

\bi

\item $\tilde{U}(4/3)$ decay:

We now consider the following possibility:
$M(\tilde{U}(4/3)) > m(\psi^{i}_{d}) + M_{\tilde{W}_{\mu\,L}}$. This
case is more or less obvious since we require the unconventional
fermions to be heavy, i.e. around $250\gev$ or so.

In this case, the dominant decay mode would be the semi-weak process:

\be
\label{decay1}
\tilde{U}(4/3) \rightarrow \bar{\psi}^{i}_{d} + \tilde{W}^{+}_{\mu\,L} \,.
\ee

Since, by assumption, $m^{2}_{d,s,b} / M^{2}_{\tilde{U}} \ll 1$, one
can immediately find the following decay width:
\be
\label{width1}
\Gamma = (\frac{G_F M^{3}_{\tilde{U}}}{8 \pi \sqrt{2}})\,
|{\cal{O}}(v^2/\delta^2)|^2 \,
|V_{\tilde{U}\,\psi^{i}_{d}}|^2\,(1-\frac{M^{2}_{\tilde{W}_{\mu\,L}}}
{ M^{2}_{\tilde{U}}})^{2}\,(1+ \frac{2\,M^{2}_{\tilde{W}_{\mu\,L}}}
{ M^{2}_{\tilde{U}}}), \,
\ee
where $G_F$ is the Fermi constant.
For $M_{\tilde{U}} \approx 250\,GeV$, one obtains

\be
\label{width1p}
\Gamma_{\tilde{U}} \approx 6\,|V_{\tilde{U}\,\psi^{i}_{d}}|^2 \,
|{\cal{O}}(v^2/\delta^2)|^2\, \gev \,,
\ee
where $V_{\tilde{U}\,\psi^{i}_{d}}$ is the matrix element of 
$V = U^{-1}_{D} U_{\tilde{U}}$. The matrices $U_{D}$ and
$U_{\tilde{U}}$ are those that diagonalize the mass matrices of
the conventional Down-quark sector and the unconventional Up-quark
sector, respectively.

The mean lifetime is found to be
\be
\label{life1}
\tau_{\tilde{U}} \approx 1.1 \times 10^{-25} |V_{\tilde{U}\,\psi^{i}_{d}}|^{-2}\,
|{\cal{O}}(v^2/\delta^2)|^{-2}\, s \,.
\ee

For illustration, let us put ${\cal{O}}(v^2/\delta^2) \approx
10^{-3}$ as we have mentioned above. The mean lifetime is
$\tau_{\tilde{U}} \approx 1.1 \times 10^{-19} |V_{\tilde{U}\,
\psi^{i}_{d}}|^{-2}\,s$.
One can see that $\tilde{U}$ can decay
{\em very fast} unless the mixing $V_{\tilde{U}\,\psi^{i}_{d}}$
is abnormally small.

For a particle which decays that fast, there is {\em no cosmological constraint}.
It can be searched for at future facilities such as the Large Hadron
Collider. This type of search was discussed at length in \cite{FHS}.

\item $\tilde{l}_{u}(-1)$ decay:

The computation for the decay rate here is very similar to that presented
above. We will assume that the mass of $\tilde{l}_{u}(-1)$ is comparable
to that of $\tilde{U}$. The main decay mode is then

\be
\label{decay2}
\tilde{l}_{u}(-1) \rightarrow \psi^{i}_{\nu} + \tilde{W}^{-}_{\mu\,L} \,,
\ee
where $\psi^{i}_{\nu} = \nu_{e}, \nu_{\mu}, \nu_{\tau}$.

Once again we will assume that $M_{\tilde{l}_{u}} \approx 250\,GeV$.
The decay width is then found to be
\be
\label{width1p2}
\Gamma_{\tilde{l}_{u}} \approx 6\,|V_{\tilde{l}_{u}\,\psi^{i}_{\nu}}|^2 \,
|{\cal{O}}(v^2/\delta^2)|^2\, \gev \,.
\ee

Its mean lifetime could be very short, i.e.
$\tau_{\tilde{l}_{u}} \approx 1.1 \times 10^{-19} 
|V_{\tilde{l}_{u}\,\psi^{i}_{\nu}}|^{-2}\,s$, if
$|V_{\tilde{l}_{u}\,\psi^{i}_{\nu}}|^2$ is not abnormally small.
Again there is {\em no cosmological constraint}. A  discussion
of the search for leptons of this type can be found in \cite{FHS}.

\ei

As we have mentioned above, although we have chosen $\tilde{U}$
and $\tilde{l}_{u}$ to illustrate how an unconventional fermion
can decay entirely into conventional particles, one can choose
other unconventional fermions to be the lightest ones and study
their decays. This will be presented elsewhere. The main point in this
section was to show that, because of mixing among the various
gauge bosons, the lightest unconventional quark or lepton is fairly
unstable, and, for the range of masses that we consider, decays
mainly into a real $W$ and a conventional fermion.

\section{Summary}
\setcounter{equation}{0}
In this paper we have extended the discussion of the early unification of
quarks and leptons based on the gauge group 
$SU(4)_{\rm PS} \otimes SU(2)_{L} \otimes
SU(2)_{R} \otimes SU(2)_{H}$ \cite{BH03}.
In particular
\begin{itemize}
\item
we have presented the Higgs system which accomplishes the spontaneous 
symmetry breaking (SSB) of the gauge group 
$G_{\rm PUT}$ down to the $SU(3)_c\otimes U(1)_{\rm QED}$ group with 
the acceptable spectrum of gauge boson, fermion and Higgs masses.
\item
we have shown that the inclusion of NLO effects and of
Higgs scalars into the renormalization group analysis increases 
the unification scale $M$,  relatively to the 
estimates in \cite{BH03}, by roughly $250\gev$ and $450\gev$ for 
$M=\tilde M$ and $M>\tilde M=800\gev$, respectively.
This  allows still for a unification of quarks and leptons
at scales $\ord(1-2~{\rm TeV})$. Specifically in the two scenario considered 
we find
\be\label{RES1F}
M=(1253\pm 132)\gev, \qquad (M=\tilde M)
\ee
and
\be\label{RES2F}
M=(1713\pm 302)\gev, \qquad (M>\tilde M\ge 800\gev).
\ee
\item
we have shown that the presence of three new generations of heavy 
unconventional quarks and leptons with masses $\ord(250\gev)$ is consistent 
with constraints coming from cosmology.
\end{itemize}
A detailed discussion of the rare decay $K_L\to\mu e$ and of FCNC processes 
will be presented elsewhere.

\vspace*{0.5truecm}

\noindent
{\bf Acknowledgements}\\
\noindent
The work presented here was supported in part by  
DFG Project Bu.\ 706/1-2.
P.Q.H. and N.K.T. are supported by the US Department of Energy
under Grant No. DE-A505-89ER40518.

\begin{appendix}
\section{Appendix A}
\setcounter{equation}{0}

In this Appendix we present the mixing of gauge bosons following the 
symmetry breaking of $U(1)_S \otimes SU(2)_L\otimes SU(2)_R\otimes SU(2)_H
\rightarrow U(1)_Y \otimes SU(2)_L\rightarrow U(1)_{QED}$.
\subsection{Neutral gauge boson mixing}
The mixing of neutral gauge bosons $W_{3R}$, $W_{3H}$, 
$\tilde{A}_{S}$ is given by the VEV of $\Delta_R(4,1,3,1)$ and 
$\Delta_H(4,1,1,3)$ (considering only color singlet part)
\be
\langle\Delta_R\rangle =
\left(\begin{array}{cc}
0 & \delta_R/\sqrt{2}  \\           
0 & 0
\end{array}\right),
\ee
\be
\langle\Delta_H\rangle =
\left(\begin{array}{cc}
0 & \delta_H/\sqrt{2}  \\           
0 & 0
\end{array}\right) 
\ee
with the corresponding covariant derivatives 
\be
D_{\mu}\Delta_R = \partial_{\mu}\Delta_R - 
i\sqrt{3/8}\;\tilde{g}_S\tilde{A}_{\mu S}(-1)\Delta_R
-ig_W[W^a_{\mu R}\frac{\tau^a}{2},\Delta_R],
\ee
\be
D_{\mu}\Delta_H = \partial_{\mu}\Delta_H - 
i\sqrt{3/8}\;\tilde{g}_S\tilde{A}_{\mu S}(-1)\Delta_H
-ig_W[W^a_{\mu H}\frac{\tau^a}{2},\delta_H]
\ee
and
\be
W^a_{\mu}\frac{\tau^a}{2}=\frac{1}{2}
\left(\begin{array}{cc}
W_{\mu 3} & \sqrt{2}W_{\mu}^{+}  \\           
\sqrt{2}W_{\mu}^{-} & -W_{\mu 3}
\end{array}\right).
\ee

From $Tr\left( D_{\mu}\Delta_R^{\dagger}D^{\mu}\Delta_R \right)$ and 
$Tr\left( D_{\mu}\Delta_H^{\dagger}D^{\mu}\Delta_H \right)$ follows the 
squared mass matrix of neutral gauge bosons

\be
\left(\begin{array}{ccc}
g_W^2\delta_R^2 & 0 & -\sqrt{3/8}\;g_W\tilde{g}_S\delta_R^2  \\           
0 &g_W^2\delta_H^2& -\sqrt{3/8}\;g_W\tilde{g}_S\delta_H^2    \\
-\sqrt{3/8}\;g_W\tilde{g}_S\delta_R^2&-\sqrt{3/8}\;g_W\tilde{g}_S\delta_H^2&
\frac{3}{8}g_S^2(\delta_R^2 +\delta_H^2)
\end{array}\right).
\ee
This matrix can be diagonalized by an orthogonal matrix $O$:
\be
O\left(\begin{array}{ccc}
g_W^2\delta_R^2 & 0 & -\sqrt{3/8}\;g_W\tilde{g}_S\delta_R^2  \\           
0 &g_W^2\delta_H^2& -\sqrt{3/8}\;g_W\tilde{g}_S\delta_H^2    \\
-\sqrt{3/8}\;g_W\tilde{g}_S\delta_R^2&-\sqrt{3/8}\;g_W\tilde{g}_S\delta_H^2&
\frac{3}{8}\tilde{g}_S^2(\delta_R^2 +\delta_H^2)
\end{array}\right) O^T =
\left(\begin{array}{ccc}
0&0&0\\
0&M_{Z_1}^2&0\\
0&0&M_{Z_2}^2
\end{array}\right)
\ee
with
\be
\left(\begin{array}{c}
B \\ Z_1 \\ Z_2
\end{array}\right)= 
O\left( \begin{array}{c}
W_{3R} \\ W_{3H} \\ \tilde{A}_s
\end{array} \right).
\ee
This diagonalization gives the following general result:

1) massless (normalized) eigenvector
\be
\left( \frac{\sqrt{3}\tilde{g}_S}{\sqrt{6\tilde{g}_S^2+8g_W^2}};
\frac{\sqrt{3}\tilde{g}_S}{\sqrt{6\tilde{g}_S^2+8g_W^2}};
\frac{2\sqrt{2}g_W}{\sqrt{6\tilde{g}_S^2+8g_W^2}} \right).
\ee

2) 1st massive neutral boson 
\be
M_{Z1}^2= \frac{1}{2} \left[(\delta_R^2 +\delta_H^2)(g_W^2+3\tilde{g}_S^2/8)-
\sqrt{(\delta_R^2 +\delta_H^2)^2(g_W^2+3\tilde{g}_S^2/8)^2-
g_W^2\delta_R^2\delta_H^2(4g_W^2+3\tilde{g}_S^2)}\right]
\ee
with corresponding (not-yet-normalized) eigenvector
\beqa
\nonu
\left(-\frac{g_W\sqrt{8}\left[(\delta_H^2 -\delta_R^2)(g_W^2+3\tilde{g}_S^2/8) +  
\sqrt{(\delta_R^2 +\delta_H^2)^2(g_W^2+3\tilde{g}_S^2/8)^2-
g_W^2\delta_R^2\delta_H^2(4g_W^2+3\tilde{g}_S^2)}
\right]}{\tilde{g}_S\sqrt{3}\left[g_W^2(\delta_H^2 -\delta_R^2)-
3\tilde{g}_S^2(\delta_R^2 +\delta_H^2)/8+
\sqrt{(\delta_R^2 +\delta_H^2)^2(g_W^2+3\tilde{g}_S^2/8)^2-
g_W^2\delta_R^2\delta_H^2(4g_W^2+3\tilde{g}_S^2)}
\right]}; \right. \\ \nonu
\left.
\frac{\sqrt{3/2}\;g_W\tilde{g}_S\delta_H^2}
{g_W^2(\delta_H^2 -\delta_R^2)-3\tilde{g}_S^2(\delta_R^2 +\delta_H^2)/8+
\sqrt{(\delta_R^2 +\delta_H^2)^2(g_W^2+3\tilde{g}_S^2/8)^2-
g_W^2\delta_R^2\delta_H^2(4g_W^2+3\tilde{g}_S^2)}}; 1 \right).
\eeqa

3) 2nd massive neutral boson 
\be
M_{Z2}^2=\frac{1}{2} \left[(\delta_R^2 +\delta_H^2)(g_W^2+3\tilde{g}_S^2/8)+
\sqrt{(\delta_R^2 +\delta_H^2)^2(g_W^2+3\tilde{g}_S^2/8)^2-
g_W^2\delta_R^2\delta_H^2(4g_W^2+3\tilde{g}_S^2)} \right]
\ee
with corresponding (not-yet-normalized) eigenvector
\beqa
\nonu
\left(-\frac{g_W\sqrt{8}\left[(\delta_H^2 -\delta_R^2)(g_W^2+3\tilde{g}_S^2/8) -  
\sqrt{(\delta_R^2 +\delta_H^2)^2(g_W^2+3\tilde{g}_S^2/8)^2-
g_W^2\delta_R^2\delta_H^2(4g_W^2+3\tilde{g}_S^2)}
\right]}{\tilde{g}_S\sqrt{3}\left[g_W^2(\delta_H^2 -\delta_R^2)-
3\tilde{g}_S^2(\delta_R^2 +\delta_H^2)/8 -
\sqrt{(\delta_R^2 +\delta_H^2)^2(g_W^2+3\tilde{g}_S^2/8)^2-
g_W^2\delta_R^2\delta_H^2(4g_W^2+3\tilde{g}_S^2)}
\right]}; \right. \\ \nonu
\left.
\frac{\sqrt{3/2}\;g_W\tilde{g}_S\delta_H^2}
{g_W^2(\delta_H^2 -\delta_R^2)-3\tilde{g}_S^2(\delta_R^2 +\delta_H^2)/8 -
\sqrt{(\delta_R^2 +\delta_H^2)^2(g_W^2+3\tilde{g}_S^2/8)^2-
g_W^2\delta_R^2\delta_H^2(4g_W^2+3\tilde{g}_S^2)}}; 1 \right).
\eeqa
A special situation where $\delta_R^2 = \delta_H^2$ has been also 
discussed in the main text.
\subsection{Charged gauge boson mixing}
The mixing of charged gauge bosons $W_{L}^{\pm}$, $W_{R}^{\pm}$, 
$W_{H}^{\pm}$ is given by the VEV of $\Delta_R(4,1,3,1)$, 
$\Delta_H(4,1,1,3)$ (see above), $\Phi(1,2,2,1)$, 
$\tilde{\Phi}_1(1,2,1,2)$, $\tilde{\Phi}_2(1,1,2,2)$, $\phi_s(15,2,2,1)$, 
$\Phi_H(1,2,2,3)$:
\bi
\It 
\be
\langle\Phi\rangle=\left(\begin{array}{cc}
v_1/\sqrt{2} & 0  \\
0 & v_2/\sqrt{2}
\end{array}\right),
\ee
\It 
\be
\langle\tilde{\Phi}_1\rangle=\left(\begin{array}{cc}
u_1/\sqrt{2} & 0  \\           
0 & u_2/\sqrt{2}
\end{array}\right),
\ee
\It 
\be
\langle\tilde{\Phi}_2\rangle=\left(\begin{array}{cc}
w_1/\sqrt{2} & 0  \\           
0 & w_2/\sqrt{2}
\end{array}\right),
\ee
\It 
\be
\langle\phi_s\rangle=\left(\begin{array}{cc}
v'_1 & 0  \\           
0 & v'_2
\end{array}\right) \otimes \frac{1}{2\sqrt{6}}
\left(\begin{array}{cccc}
1 & 0&0&0  \\           
0 & 1&0&0  \\
0 & 0&1&0  \\
0 & 0&0&-3
\end{array}\right),
\ee
\It
\be
\langle\Phi_H\rangle=\left(\begin{array}{cccc}
\chi_1/2 & 0 & 0& 0  \\           
0 & \chi_2/2 & 0 & 0  \\
0 & 0 & -\chi_1/2 & 0  \\
0 & 0& 0&  -\chi_2/2
\end{array}\right),
\ee
\ei
with the corresponding covariant derivative 
\be
D_{\mu}\Phi = \partial_{\mu}\Phi 
-ig_W[W^a_{\mu L}\frac{\tau^a}{2}]\Phi
-ig_W\Phi[W^a_{\mu R}\frac{\tau^a}{2}],
\ee
\be
D_{\mu}\tilde{\Phi}_1 = \partial_{\mu}\tilde{\Phi}_1 
-ig_W[W^a_{\mu L}\frac{\tau^a}{2}]\tilde{\Phi}_1
-ig_W\tilde{\Phi}_1[W^a_{\mu H}\frac{\tau^a}{2}],
\ee
\be
D_{\mu}\tilde{\Phi}_2 = \partial_{\mu}\tilde{\Phi}_2 
-ig_W[W^a_{\mu R}\frac{\tau^a}{2}]\tilde{\Phi}_2
-ig_W\tilde{\Phi}_2[W^a_{\mu H}\frac{\tau^a}{2}],
\ee
\be
D_{\mu}\phi_s \supset \partial_{\mu}\phi_s 
-ig_W[W^a_{\mu L}\frac{\tau^a}{2}] \otimes \phi_s
-ig_W\phi_s\otimes[W^a_{\mu R}\frac{\tau^a}{2}],
\ee
\be
D_{\mu}\phi_{223} = \partial_{\mu}\phi_{223}
-ig_W[W^a_{L\mu}\frac{\tau^a}{2}] \otimes \phi_{223}
-ig_W\phi_{223}\otimes [W^a_{R\mu}\frac{\tau^a}{2}] 
-ig_W[W^a_{H\mu}\frac{\tau^a}{2},\phi_{223}].
\ee

From $Tr\left( D_{\mu}\phi^{\dagger}D^{\mu}\phi \right)$ follows the 
squared mass matrix of charged gauge bosons $W_{L}^{\pm}$, $W_{R}^{\pm}$, 
$W_{H}^{\pm}$.
\be
g_W^2\left(\begin{array}{ccc}
V^2+u^2+\chi^2 & V_{12}^2+\chi_{12}^2 
&u_{12}^2\\  
V_{12}^2+\chi_{12}^2 &\delta_R^2/2+V^2+w^2+\chi^2  
&w_{12}^2\\
u_{12}^2 &w_{12}^2 & \delta_H^2/2 +u^2 +w^2+4\chi^2
\end{array}\right) 
\ee
where we have defined
\be
V^2= \frac{v_1^2 + v_2^2}{4}+\frac{{v'}_1^2 + {v'}_2^2}{4}; \;\;\;\;\;
u^2= \frac{u_1^2 + u_2^2}{4}; \;\;\;\;\;
w^2= \frac{w_1^2 + w_2^2}{4};
\ee
\be
\chi^2=\frac{\chi_1^2+\chi_2^2}{4};
\ee
\be
\chi_{i,j}^2=\frac{\chi_i \chi_j}{2}; \;\;\;\;\;
V_{i,j}^2=\frac{v_i v_j}{2}+\frac{{v'}_i {v'}_j}{2}; \;\;\;\;\;
u_{i,j}^2=\frac{u_i u_j}{2}; \;\;\;\;\;
w_{i,j}^2=\frac{w_i w_j}{2}.
\ee
In the limit $ v_i^2,{v'}_m^2, 
u_k^2, \chi_j^2 \ll w_l^2, \delta_R^2, \delta_H^2$ (as it is considered in 
the main text), we obtain in the leading order approximation
\bi
\It 1st massive charged boson's squared mass
\be
M_{\tilde{W}_L}^2=g_W^2\left(V^2 +u^2 +\chi^2
+{\cal{O}}(\frac{v^4}{\delta^2}) \right)
\ee 
with the corresponding (normalized) eigenvector
\be
\left(1+{\cal{O}}(\frac{v^2}{\delta^2})\;\;;\;\; 
{\cal{O}}(\frac{v^2}{\delta^2})\;\;;\;\; 
{\cal{O}}(\frac{v^2}{\delta^2}) \right).
\ee

\It 2nd massive charged boson's squared mass
\be
M_{W_1}^2=g_W^2\left(\frac{\delta_R^2+\delta_H^2}{4} +
\frac{1}{4}\sqrt{(\delta_R^2-\delta_H^2)^2+16w_{12}^4}
+{\cal{O}}(v^2) \right)
\ee 
with the corresponding (normalized) eigenvector


\be
\frac{\left[
{\cal{O}}(v^2) \;;\; 2\sqrt{2}w_{12}^2 + 
{\cal{O}}(v^2) \;;\; 
\frac{\delta_H^2-\delta_R^2 +\sqrt{(\delta_R^2-\delta_H^2)^2+
16w_{12}^4}}{\sqrt{2}}\,+\,{\cal{O}}(v^2) \right]}
{\sqrt{16w_{12}^4 +
(\delta_R^2-\delta_H^2)^2 -(\delta_R^2-\delta_H^2)\sqrt{(\delta_R^2-\delta_H^2)^2+16w_{12}^4}}}.
\ee

\It 3rd massive charged boson's squared mass
\be
M_{W_2}^2=g_W^2\left(\frac{\delta_R^2+\delta_H^2}{4} -
\frac{1}{4}\sqrt{(\delta_R^2-\delta_H^2)^2+16w_{12}^4}
+{\cal{O}}(v^2) \right)
\ee 
with the corresponding (normalized) eigenvector
\be
\frac{\left[{\cal{O}}(v^2) \; ; \;
2\sqrt{2}w_{12}^2 +{\cal{O}}(v^2)\; ; \; 
\frac{\delta_H^2-\delta_R^2 -\sqrt{(\delta_R^2-\delta_H^2)^2 +
16w_{12}^4}}{\sqrt{2}} \,+\,{\cal{O}}(v^2) \right]}
{\sqrt{16w_{12}^4 +
(\delta_R^2-\delta_H^2)^2 +
(\delta_R^2-\delta_H^2)\sqrt{(\delta_R^2-\delta_H^2)^2+16w_{12}^4}}}.
\ee
\ei
Having these eigenstates, one can straightforwardly construct the orthogonal matrix $O$ that 
diagonalizes the squared mass matrix of charged bosons. The same matrix also relates 
the gauge eigenstates $W_L^{\pm}$, $W_R^{\pm}$, $W_H^{\pm}$
to the mass eigenstates $\tilde{W}_L^{\pm}$, $W_1^{\pm}$, $W_2^{\pm}$ 
\be
\left(\begin{array}{c}
W_L^{\pm} \\ W_R^{\pm} \\ W_H^{\pm}
\end{array}\right)= 
O^T\left( \begin{array}{c}
\tilde{W}_{L}^{\pm} \\W_{1}^{\pm} \\ W_{2}^{\pm}
\end{array} \right).
\ee
In a special case where $\delta_R^2 = \delta_H^2 \equiv \delta^2$, and in the leading order the above expressions become quite simple:
\bi
\It 1st massive charged boson's squared mass
\be
M_{\tilde{W}_L}^2=g_W^2\left(V^2 +u^2 +\chi^2
+{\cal{O}}(\frac{v^4}{\delta^2}) \right)
\ee 
with the corresponding (normalized) eigenvector
\be
\left(1 + {\cal{O}}(\frac{v^2}{\delta^2}) \;;\; 
{\cal{O}}(\frac{v^2}{\delta^2})\;;\;  
{\cal{O}}(\frac{v^2}{\delta^2})\right).
\ee

\It 2nd massive charged boson's squared mass
\be
M_{W_1}^2=g_W^2\left(\frac{\delta^2}{2} +w_{12}^2 +{\cal{O}}(v^2) \right)
\ee 
with the corresponding (normalized) eigenvector
\be
\left({\cal{O}}(\frac{v^2}{\delta^2})\;;\; \frac{1}{\sqrt{2}} \,+\,
{\cal{O}}(\frac{v^2}{\delta^2})\;;\; \frac{1}{\sqrt{2}} \,+\,
{\cal{O}}(\frac{v^2}{\delta^2}) \right).
\ee

\It 3nd massive charged boson's squared mass
\be
M_{W_2}^2=g_W^2\left(\frac{\delta^2}{2} - w_{12}^2 +{\cal{O}}(v^2) \right)
\ee 
with the corresponding (normalized) eigenvector
\be
\left({\cal{O}}(\frac{v^2}{\delta^2})\;;\; \frac{1}{\sqrt{2}} \,+\,
{\cal{O}}(\frac{v^2}{\delta^2})\;;\; -\frac{1}{\sqrt{2}} \,+\,
{\cal{O}}(\frac{v^2}{\delta^2})\right).
\ee

\ei
 From these eigenstates, one can construct the rotation matrix, 
\be
\left(\begin{array}{c}
W_L^{\pm} \\ W_R^{\pm} \\ W_H^{\pm}
\end{array}\right)= 
O^T\left( \begin{array}{c}
{\tilde{W}}_{L}^{\pm} \\W_{1}^{\pm} \\ W_{2}^{\pm}
\end{array} \right) 
\ee
which has the following form:
\be
O^T = \left(\begin{array}{ccc}
1& {\cal{O}}(\frac{v^2}{\delta^2})  & {\cal{O}}(\frac{v^2}{\delta^2}) \\
{\cal{O}}(\frac{v^2}{\delta^2})& 1/\sqrt{2}  & 1/\sqrt{2} \\
{\cal{O}}(\frac{v^2}{\delta^2})& 1/\sqrt{2}  & -1/\sqrt{2}
\end{array} \right).
\ee

\end{appendix}
\end{document}